\documentclass[%
 aip,
jcp,%
 amsmath,amssymb,
 preprint,%
]{revtex4-1}
\UseRawInputEncoding
\usepackage{graphicx}% Include figure files
\usepackage{dcolumn}% Align table columns on decimal point
\usepackage{bm}% bold math
\usepackage{xcolor}

\usepackage{amsmath}
\usepackage{amsfonts}
\usepackage{amssymb}
\usepackage{graphicx}
\usepackage{hyperref}
\usepackage[version=3]{mhchem} % Formula subscripts using \ce{}
\usepackage[T1]{fontenc} 
\usepackage{simplewick}
\usepackage{wick}
\usepackage{diagbox}
\usepackage{alltt}
\usepackage{braket}
\usepackage{float}
\usepackage[]{cancel}
\begin{document}

\preprint{AIP/123-QED}

\title[]{Exploration of interlacing and avoided crossings in a manifold of potential energy curves by a Unitary Group Adapted State Specific Multi-Reference Perturbation Theory (UGA-SSMRPT)}

\author{Dibyajyoti Chakravarti}
\affiliation%[IACS]
{School of Chemical Sciences, Indian Association for the Cultivation of Science, India}
\email{rana.chakravarti@gmail.com}
\author{Koustav Hazra}
\affiliation%[IACS]
{School of Chemical Sciences, Indian Association for the Cultivation of Science, India}
\email{koustav.hazra93@gmail.com}
\author{Riya Kayal}
\affiliation%[IACS]
{School of Chemical Sciences, Indian Association for the Cultivation of Science, India}
\email{riyakayal.rk@gmail.com}
\author{Sudip Sasmal}
\affiliation%[Heidelberg]
{Physikalisch-Chemisches Institut, Universit\"{a}t Heidelberg, Germany}
\email{sudip.sasmal@pci.uni-heidelberg.de}
\author{Debashis Mukherjee}
\affiliation%[SNBNCBS]
{Centre for Quantum Engineering, Research, and Education (CQuERE),
TCG-CREST,India}
\email{pcdemu@gmail.com}

\date{\today}

\begin{abstract}
%Any effective Hamiltonian based many-body formalism describing a manifold of states dominated by a set of ‘model functions’ spanning a CAS is numerically unstable due to the notorious problem of intruders. This is why generating PES of a group of states can rarely be studied in a size-consistent manner via an H$_{eff}$ . CAS based state-specific theories which target just one root of interest at a time has the potentiality of bypassing intruders, yet providing accurate PES. 
The Unitary Group Adapted State-Specific Multi-Reference Perturbation Theory (UGA-SSMRPT2) developed by Mukherjee et al [J. Comput. Chem. 2015, 36, 670 ; J. Chem. Theory Comput. 2015, 11, 4129] has successfully realized the goal of studying bond dissociation in a numerically stable, spin-preserving and size-consistent manner. In this paper, we explore and analyse the UGA-SSMRPT2 theory in the description of avoided crossings and interlacing between a manifold of states belonging to the same space-spin symmetry. It is not immediately obvious how a state-specific theory, generating successively higher-lying PEC one at a time, would retain sufficiently accurate information of other close lying PEC of the same symmetry. In a state-specific formalism, since each state is an eigenstate of its own effective operator, to include the information of the other states requires the theory to be sufficiently accurate. Three different aspects of UGA-SSMRPT2 have been studied: (a) We introduce and develop the most rigorous version of UGA-SSMRPT2 which emerges from the rigorous version of UGA-SSMRCC utilizing a linearly independent virtual manifold; we call this the 'projection' version of UGA-SSMRPT2 denoted as UGA-SSMRPT2 Scheme P. We compare and contrast this approach with our earlier formulation that used extra sufficiency conditions via amplitude equations, which we will denote as UGA-SSMRPT2 Scheme A. (b) We present the results for a variety of electronic states of a set of molecules which display the striking accuracy of both the two versions of UGA-SSMRPT2; with respect to three different situations involving weakly avoided crossings, moderate/strongly avoided crossings and interlacing in a manifold of PECs of same symmetry. Accuracy of our results has been benchmarked against IC-MRCISD+Q. (c) For weakly avoided crossing between states displaying differently charged sectors in the asymptotes, the insufficient inclusion of state-specific orbital relaxation in a second order perturbative theory might lead to an artefact of double crossing between the pair of PECs.
\end{abstract}

\maketitle
\section{Introduction} \label{Intro}
Multireference (MR) electron correlation theories have been a rapidly emerging area of research in recent years \cite{Lyakh2012MultireferenceView,Szalay2012MulticonfigurationApplications,Evangelista2018Perspective:Correlation} for the proper description of excited state behaviour and dissociation profile of chemically relevant molecules. Such theories by design are formulated in a given N-electron sector of the Fock space, referred to as the Hilbert space, and are thus structurally quite different from the earlier developed multireference correlation theories which used valence universal wave operators\cite{Mukherjee1975CorrelationMolecules,Mukherjee1977ApplicationsTrans-butadiene,Lindgren1978ASystems,Haque1985Open-shellSystems,Lindgren1987OnSpaces}. For most of the MR methods, a generic strategy is to divide the electron correlation into non-dynamical, which is a strong correlation resulting from near-degeneracy of certain configurations, and dynamical correlation, which results from the configurations which are excited with respect to the near-degenerate configurations (called the active space). The non-dynamical or static correlation, being the stronger, needs a more accurate description which is introduced commonly via a full configuration interaction of all the quasi-degenerate configurations, called the complete active space (CAS).
The dynamical correlation energy contribution is usually an order of magnitude less but matches the order of chemical accuracy required in a spectroscopic or chemical phenomenon. It can be computed using either a non-perturbative or a perturbative approach, and it is necessary to properly theorize its contribution with certain desirable properties like (a) size extensivity \cite{Bartlett1981Many-BodyMolecules}, (b) size consistency \cite{Pople1978ElectronSurfaces} and (c) invariance with respect to the orbital basis used.
\par
Dynamical correlation effects are usually incorporated using various strategies: MR variants of CI theory \cite{Werner1988AnIN}, coupled cluster (CC) theories\cite{Jeziorski1981Coupled-clusterStates,Li1995UnitaryApproximations,Li1997ReducedStates,MAHAPATRA1998AApplications,Pittner1999AssessmentMethod:TestcalculationsonCH2SiH2andtwistedethylene,Hanrath2005AnAnsatz,Kong2009StateSpace,Herrmann2020GenerationOrder,Morchen2020TailoredRegimes} including the unitary versions thereof \cite{Yanai2006CanonicalProblems,Chen2012OrbitallyApproximation}, perturbation theories (PT) \cite{Andersson1992SecondorderFunction,Hirao1992MultireferenceMethod,Kozlowski1994ConsiderationsTheory,Hoffmann1996CanonicalVariables,Angeli2001IntroductionTheory,Chaudhuri2005ComparisonTheories,Fink2009TheApproach,Mao2012ADevelopments,Liu2014SDS:Electrons,Sharma2014Communication:States,Sen2015UnitaryApplications,Li2015MultireferenceAnalysis,Zhang2020IterativeSelection} and other allied approaches \cite{Booth2009,Thom2010,Schollwock2010,Chan2011a,Knecht2016,Holmes2016Heat-BathSampling,Deustua2017,Scott2019,Filip2019,Baiardi2020}. The general problem with MRCI is that, despite using a complete active space, the inclusion of virtual functions which are excited with respect to the CAS functions upto a given rank lead in general to size inextensive energies, although empirical corrections to alleviate the severity of this error have been in use \cite{Szalay1995}. Most of the CC and PT formulations involve construction of an effective Hamiltonian (H$_{eff}$) in the active space, which contains the effect of both kinds of correlation folded into it via a wave operator ($\Omega$). The diagonalization of this H$_{eff}$ in the active space gives us the energies of the states of interest.
\par
More often than not, use of a complete active space to generate H$_{eff}$, although seemingly elegant in structure, is seriously affected by numerical instability due to the notorious and ubiquitous 'intruder state' problem \cite{Schucan1973}.
We do not want to go into the details of the intruder problem here, as the practitioners of multireference correlation theories are well aware of its origin and deleterious effects.
\par
Several approaches have been proposed to ameliorate the intruder problem which differ considerably among themselves both in their approach and in detail. They can be broadly classified into three categories which differ in their treatment of the dynamical correlation: (i) an effective Hamiltonian in a suitably chosen incomplete active space \cite{Mukherjee1986TheSpaces,Mukherjee1986AspectsTheory} (ii) an Intermediate Hamiltonian approach \cite{Malrieu1985} and (iii) state-selective (targeting specific states of interest) approach \cite{MAHAPATRA1998AApplications,Pittner1999AssessmentMethod:TestcalculationsonCH2SiH2andtwistedethylene,Hanrath2005AnAnsatz}.
\par
The incomplete model space approach involves the selection of an appropriate subspace of the CAS in defining effective Hamiltonians such that the active space functions which strongly interact with the intruding virtual functions are pushed into the virtual manifold, and then diagonalizing the H$_{eff}$ in this truncated space. It is very difficult to maintain size extensivity for the corresponding energies evaluated by this approach due to the incompleteness of the model space, but a rigorously size-extensive approach was proposed by Mukherjee which involves abandoning the intermediate normalization condition (IN) for the eigenfunctions \cite{Mukherjee1986TheSpaces,Mukherjee1986AspectsTheory}.
\par
The intermediate Hamiltonian formulation is an ingenious manipulation where the effective Hamiltonian is not truncated but it is divided into two subsets, one containing the states of interest and the other the states which are intruder prone (the intermediate space). A level shift is given to the solution of the intermediate states which numerically removes their divergence and a condition for vanishing the coupling between the intermediate and the desired states is imposed. This manipulation leads to numerical stability for the states of interest despite diagonalizing in the complete model space. The only limitation of this approach is the lack of size extensivity of the energies computed, unless explicit constraints are imposed to ensure size extensivity \cite{Mukherjee1986AspectsTheory,Mukhopadhyay1992TheFramework,Malrieu2013ProposalFormalisms}. An extreme approach to remove the intruder problem is to use the effective Hamiltonian approach but solve for cluster amplitudes for one specific state of interest at a time. This has been the most promising approach among the three, but a formulation which satisfies all the desirable attributes (a-c as mentioned earlier)  for the evaluated energy remains a challenge.
\par
Looked at from a very accurate inclusion of both dynamical and non-dynamical correlation, the state-specific version of MRCC developed by Mukherjee et al\cite{MAHAPATRA1998AApplications,Maitra2012UnitaryApplications,Evangelista2018Perspective:Correlation}, using a decontracted wave operator ansatz of the Jeziorski-Monkhorst type \cite{Jeziorski1981Coupled-clusterStates}, has been shown to have both the desirable properties of size extensivity and intruder-free solutions, but lacks orbital invariance in the energy and other properties computed. On the other hand, an internally contracted formulation (IC-MRCC) \cite{Banerjee1981TheState,Mukherjee1995AState,Hanauer2011PilotProperly} suffers no orbital invariance problem, but it is at the expense of a huge redundancy problem in the virtual manifold, which has to be removed via a computationally expensive extraction of an orthogonal excitation space.\par
On the other end of the spectrum, a PT approach to dynamic correlation would have been much cheaper and hence more attractive for applications to larger chemically relevant systems, and has garnered a lot of attention in the recent past. The error in the absolute energy computed by a PT would be far less relevant in comparison to the correct explanation of trends in the studied chemical phenomenon. The earliest developments were based on perturbative approximations to MRCISD, and two of the most popular of these are the CASPT2 and the MRMP2 methods developed by Roos et al\cite{Andersson1992SecondorderFunction} and Hirao \cite{Hirao1992MultireferenceMethod} respectively. Both of them are second order SS theories and utilize a contracted unperturbed function. CASPT2 suffers from substantial errors in size extensivity and consistency, along with a notorious problem of numerical instability due to the necessity of a larger CAS in the theory than chemically required to describe the studied phenomenon. The numerical instability that plagues the contracted SSMRPTs is mainly due to lack of coefficient relaxation in the perturbed wavefunction and has been well discussed in a review by Malrieu et al. \cite{Malrieu1995MulticonfigurationalProperties}. The original MRMP2 theory too is size inextensive and not orbital invariant, although an improved version has been later reported by van Dam et al \cite{vanDam1999ExactTheory}.The NEVPT2 developed by Malrieu et al\cite{Angeli2001IntroductionTheory} eliminates the need of extracting the linearly independent excitation manifold by an ingenious definition of the excitation operators. It has both SS and intermediate Hamiltonian versions. Several choices of unperturbed Hamiltonian have been used, with some of them being able to minimize the intruder state problem. The fully general version of the theory is both size extensive and orbital invariant, but computationally expensive. The GVVPT2 developed by Hoffmann \cite{Hoffmann1996CanonicalVariables} is based on a unitary wave operator and is an intermediate Hamiltonian based approach. It is not rigorously size-extensive but the numerical errors have been found to be small in molecular applications. A recent development by Liu et al \cite{Liu2014SDS:Electrons} is the SDSPT2 theory which is also an intermediate Hamiltonian based method and it too is not strictly size-extensive.
\par
The UGA-SSMRPT2 theory developed by Mukherjee et al is a perturbative approximant of the UGA-SSMRCC theory developed by the group, and it preserves all the desirable qualities of being strictly size extensive and intruder-free for the ground and low-lying excited states\cite{Maitra2012UnitaryApplications,Sen2015UnitaryApplications}. The theory stands out from the other MRPTs discussed briefly earlier, owing to the fact that it is systematically improvable order by order upto a fully developed non-perturbative theory, i.e. UGA-SSMRCC. The only other MRPT which has a corresponding non-perturbative analogue is the GVVPT2 theory, which is a perturbative version of a unitary MRCC theory \cite{Hoffmann1988AApplications}. Neither the UGA-SSMRPT2 nor the UGA-SSMRCC theories are orbital invariant owing to the use of a Jeziorski-Monkhorst (J-M) like\cite{Jeziorski1981Coupled-clusterStates} decontracted wave operator.
\par
The original spinorbital-based formulation of the SSMRPT2 required the use of suitable sufficiency conditions for determining the cluster amplitudes. The need for sufficiency conditions arises because a given virtual function can be reached from more than one model function, implying more than one model-function dependent cluster amplitude for a given virtual function. In a spinorbital based formalism the virtual functions are all linearly independent. For a spin-adapted SSMRPT formulation, a virtual function CSF is not uniquely specified by the orbital occupancies alone because of the so-called spin-degeneracy problem \cite{Pauncz1979SpinUse}. This poses an additional linear dependence in the UGA-based state specific many-body theories using the J-M ansatz as is the case in the UGA-SSMRCC~\cite{Maitra2012UnitaryApplications} and UGA-SSMRPT~\cite{Sen2015UnitaryApplications} generated therefrom.
In our opinion, there are two different ways to resolve this problem: (i) To generate suitable amplitude equations we may posit suitable \textit{additional sufficiency conditions}. The resultant 'amplitude' equations will henceforth be denoted as Scheme A. A similar idea was first suggested by us in a previous spin-adapted MRPT formulation as well\cite{Mao2012ADevelopments}. (ii) One can however bypass the use of this extra sufficiency condition by choosing only those CSFs for projections which are linearly independent as demanded by the spin degeneracy. The UGA based formalisms allows one to naturally choose only those CSFs. This latter choice leads to another, inequivalent, formalism which one may call 'projection' equations (to be henceforth called Scheme P).
Both of these developments will be discussed in Section \ref{theory} of our paper, first for the UGA-SSMRCC and then for the UGA-SSMRPT2 generated therefrom.
\par
We should mention here a recent work by Giner et al \cite{Giner2017ACurves} which formulated the spin-adapted JMMRPT2 theory which also uses the Jeziorski-Monkhorst(J-M) ansatz in its wave operator and is also not orbital invariant. The size extensivity of its computed energy has been enforced using a clever manipulation of the denominator that is used to solve for the cluster amplitudes. But this novelty is probably limited to the perturbative level and does not seem to be easily generalizable to a non-perturbative theory.
\par
One might wonder about the efficacy of an SS-MR theory in either a perturbative or CC approach to describe PECs of a manifold of states displaying avoided crossings when the theory is employed for each of the states separately. In fact, this demands more flexibility and accuracy on a theory to sense the presence of neighbouring states involved in the various modes of avoided crossing. In this paper, we shall explore the effectiveness of our UGA-SSMRPT2 theory using both of it's versions (scheme A and P) in treating a manifold of states of same space-spin symmetry, particularly highlighting regions of single or multiple avoided crossings, the latter to be called 'interlacing'. Such avoided crossings demand reproduction of the rather sensitive change in gradient in that region in the dissociation profiles of each state. It is not immediately obvious that the manifold of states which are targeted using separate state-specific computations would 'sense' each other in terms of the intricate gradient changes with change in geometry and would exhibit their interlacing with each other. Moreover the positions of these avoided crossings are sensitive to the inclusion of dynamical correlation and the state-specific relaxation of active orbitals attendant on such dynamical correlation. We shall highlight the importance of state-specific orbital relaxation especially for describing the weakly avoided crossings involving dramatic differences in the relative weightage of ionic and covalent character of the states concerned. In a perturbative computation of second-order energy, the first-order perturbed function contains insufficient orbital relaxation effects which can be crucial when the relaxation is strong. Since it is natural to use a common set of orbitals to describe PECs of close-lying states, the interplay of state-specific orbital relaxation and dynamical correlation in a perturbative theory such as UGA-SSMRPT2 becomes crucial to ascertain. In this paper the formulation of scheme P is presented in some detail since it has never been discussed previously.
We compare the use of operator equations (scheme A) involving sufficiency conditions against the proper projection scheme (scheme P). It will be shown that the errors with respect to MRCI methods is comparable in both the schemes, thus suggesting the efficacy of our sufficiency which significantly reduces the computational time of our calculations.
\par
Our paper is organized as follows: Section II traces the rigorous genesis of the UGA-SSMRPT2 suggested in this paper from the rigorous UGA-SSMRCC theory. The underlying theoretical issues distinguishing the amplitude and projection schemes and the corresponding working equations are discussed next. The aspects of state-specific orbital relaxation, in particular for situations involving weakly avoided crossings, are also introduced and discussed here. Section III contains the molecular applications and relevant discussions on various PECs of a selection of manifold of states of prototypical molecules. The molecules chosen by us are such that three different aspects of PEC viz the medium and strongly avoided crossing, the weakly avoided crossing and the interlacing of various PECs in a given symmetry manifold can be demonstrated. Section IV summarizes the highlights of our findings and presents our future outlook.
\section{Theory}\label{theory}
\subsection{Description of a rigorous UGA-SSMRCC and its approximant, UGA-SSMRPT2, for second order energy}
As mentioned in Sec.~\ref{Intro}, the earlier UGA-SSMRPT2 theory was developed by Mukherjee et al manifestly in the amplitude form \cite{Sen2015UnitaryApplications,Sen2015AspectsPrototype}. It has been shown to work very well to describe complex bond dissociation profiles of N$_2$, B$_2$, C$_2$, O$_2$, etc. The low-lying excited states of these molecules with different space-spin symmetry have also been computed successfully. %This theory has also been extensively applied by Chattopadhyay et al \cite{Chattopadhyay2015State-specificN2} and it bypasses the use of CASSCF orbitals by a numerically lower-scaling orbital optimization called the Improved Virtual Orbital (IVO) \cite{Potts2001TheStates} method.
%The theory is based on the SSMRPT by Mukherjee et al \cite{SinhaMahapatra1999MolecularCoefficients} although the details of its spin-adaptation and a lucid discussion on the inevitable spin redundancy problem involved is absent from their publications.\\
The theory is an intrinsically low-scaling and a rigorously size-extensive, intruder-free, state-specific multireference theory. We summarize the main tenets of the theory that help us to arrive at the final working equations.
\par
Unlike many of the multireference perturbation theories described in the Section \ref{Intro}, UGA-SSMRPT2 is truly a second order approximant originating from a general non-perturbative correlation theory. It is derived from the UGA-SSMRCC\cite{Maitra2012UnitaryApplications} theory, preserving all of its desirable qualities such as size-extensivity and intruder-free nature, with the additional advantage of a low computational cost. The essential motivation in generating the UGA-SSMRPT2 from UGA-SSMRCC is to find an unperturbed Hamiltonian H$^{(0)}$ which captures much of the important physics at the lowest order. In the UGA-SSMRPT2 theory, multi-partitioning of H$^{(0)}$ is the natural choice, i.e. the H$^{(0)}$ varies for each model function $\phi_\mu$ in the CAS chosen. The idea of multi-partitioning of the unperturbed Hamiltonian was first introduced by Malrieu et al \cite{Zaitsevskii1995Multi-partitioningTheory,Zaitsevskii1996Multi-partitioningFormulation,Zaitsevskii1997Spin-adaptedTheory}.
\par
The parent UGA-SSMRCC introduced a wave operator $\Omega$ of the Jeziorski Monkhorst\cite{Jeziorski1981Coupled-clusterStates} (J-M) type, but posited a normal-ordered exponential cluster ansatz, essentially to exclude the cumbersome contractions between the cluster operators.
\begin{equation}\label{ansatz}
\Omega=\sum_\mu \Omega_\mu=\sum_\mu\{e^{T_\mu}\}\ket{\phi_\mu}\bra{\phi_\mu}
\end{equation}
The curly bracket in Eq.\ref{ansatz} indicates normal ordering with respect to a suitable closed-shell vacuum.
For a detailed discussion on this Ansatz and it's advantages over other spin-free theories, we refer the readers to recent literature from our group\cite{Maitra2012UnitaryApplications,Sen2012FormulationEnergies,Sen2018InclusionElements}. 
We have recently derived a rigorous and systematically improvable UGA-SSMRCC theory which is the closest spin-free analogue to the spinorbital SSMRCC theory derived by us earlier \cite{MAHAPATRA1998AApplications}. This derivation and a comprehensive discussion would be presented in a soon to be published article \cite{TBP}.
\par
For any many-body operator A$_\mu$, we can decompose it as a sum of a 'closed' operator A$_\mu^{cl}$, which when acting on a CSF $\phi_\mu$, spanning the active space, leads only to transition within the CAS functions; and an 'excitation' operator A$_\mu^{ex}$ which leads to virtual functions (orthogonal to the CAS functions) when acting on $\phi_\mu$.\\
The working equation for determining the cluster amplitudes T$_\mu$ is of the form:
\begin{equation}
    G_\mu^{ex}\ket{\phi_\mu}=0\ \ \forall\ \mu
\end{equation}
The coefficients \{c$_\nu$ $\forall\ \nu$\} are to be obtained from the equations:
\begin{equation}
    \sum_\nu G_\nu^{cl}\ket{\phi_\nu}c_\nu=E\ket{\phi_\mu}c_\mu\ \ \forall\ \mu
\end{equation}
We present below the operator form of the rigorous CC working equation\cite{TBP} for the cluster amplitudes \{T$_\mu$\}, without its detailed derivation, to indicate the genesis of the rigorous UGA-SSMRPT2 following from it.\\
\begin{eqnarray}
    \{\contraction{}{e^{\theta_\mu}}{}{\Big[\overline{H}_\mu^{ex} +\displaystyle\sum_{\nu}e^{T_\nu-T_\mu} \widetilde{H}_{\mu\nu} \frac{c_{\nu k}}{c_{\mu k}} - \displaystyle\sum_{\nu}\contraction{}{\tilde{e}^{T_\mu}}{}{W_{\nu\mu}}{\tilde{e}^{T_\mu}W_{\nu\mu}} \Big]}{e^{\theta_\mu} \Big[\overline{H}_\mu^{ex} +\displaystyle\sum_{\nu}e^{T_\nu-T_\mu} \widetilde{H}_{\mu\nu} \frac{c_{\nu k}}{c_{\mu k}} - \displaystyle\sum_{\nu}\contraction{}{\tilde{e}^{T_\mu}}{}{W_{\nu\mu}}{\tilde{e}^{T_\mu}W_{\nu\mu}} \Big]}\}|\phi_\mu\rangle = 0 \ \ \forall\ \mu \label{finalSSopeq}
\end{eqnarray}
where, $\theta_\mu$ is defined as~\cite{Sen2018InclusionElements,TBP},
\begin{equation}
\theta_\mu = -T_\mu + \contraction{}{T_\mu}{}{T_\mu}{T_\mu T_\mu} - \contraction{}{T_\mu}{T_\mu}{T_\mu}\contraction{T_\mu}{T_\mu}{}{T_\mu}{T_\mu T_\mu T_\mu} + ...
\end{equation}
The first order perturbed operator equation will follow from the linearized terms of the equation \ref{finalSSopeq} above.
\begin{equation} \label{sscepaeqn}
     \{H_{ex}+\contraction{}{H}{}{T_\mu}H T_\mu\}\ket{\phi_\mu}c_\mu-\sum_\nu\{\contraction{}{T_\mu}{}{{H}_{\nu\mu}}T_\mu {H}_{\nu\mu}\}\ket{\phi_\mu}c_\mu \\
+\sum_\nu\{T_\nu-T_\mu\}\ket{\phi_\mu}H_{\mu\nu}c_\nu=0
\end{equation}
It is pertinent to mention here that an earlier UGA-SSMRCC was developed by Maitra et al \cite{Maitra2012UnitaryApplications} whose working equations were an approximation to Eq. \ref{finalSSopeq}, where the operator $\theta_\mu$ did not appear. Interestingly enough, it turns out that the first order perturbed equation , either in amplitude or projection form, would still not involve the operator $\theta_\mu$ since it starts contributing only from 2nd order onwards. This aspect of the theoretical content of the working equations for UGA-SSMRPT2 was not noted before.
\par
The final working equation for UGA-SSMRPT2 cluster amplitudes is thus identical to the one derived in our earlier paper by Sen et al\cite{Sen2015UnitaryApplications}. It is obtained by partitioning H as H$^{(0)}$+V and collecting terms up to first order of perturbation in Eq. \ref{sscepaeqn}, as follows,
\begin{equation} \label{UGASSMRPTopeqn}
     V\ket{\phi_\mu}c^0_\mu+\{\contraction{}{H^{(0)}}{}{T^{(1)}_\mu}H^{(0)}T^{(1)}_\mu\}_{ex}\ket{\phi_\mu}c^0_\mu-\sum_\nu\{\contraction{}{T^{(1)}_\mu}{}{{H^{(0)}}_{\nu\mu}}T^{(1)}_\mu {H^{(0)}}_{\nu\mu}\}_{ex}\ket{\phi_\mu}c^0_\mu \\
+\sum_\nu\{T^{(1)}_\nu-T^{(1)}_\mu\}\ket{\phi_\mu}H_{\mu\nu}c^0_\nu=0
\end{equation}
 \subsection{Emergence of amplitude equations (Scheme A) in UGA-SSMRPT2}
 As mentioned towards the end of Section \ref{Intro}, to avoid the computationally involved procedure of extracting the linearly independent virtual space, one may posit a sufficiency condition called 'amplitude' equations. If we use this sufficiency to solve for the cluster amplitudes, we can afford to use a set of overcomplete projections onto a linearly dependent basis \{$\chi_\mu^{l^\prime}$\}; writing,
 \begin{equation}
     G_\mu^{ex}=\sum_{l^\prime}g^{l^\prime}_\mu\{E^{l^\prime}_\mu\}
 \end{equation} 
 and using the following as sufficiency conditions, we get a set of cluster amplitudes \{t$^{l^\prime}_\mu$\},
 \begin{equation}
     g^{l^\prime}_\mu=0,\ \ \forall\ l^\prime,\mu
 \end{equation}
The l$^\prime$ denotes all possible changes in orbital occupancy from a particular $\phi_\mu$.
It should be noted here that all the residue blocks (g), which correspond to a specific orbital excitation operator, have to be clubbed together using the appropriate transformation factors. This essentially means that all direct and exchange spectator scattering blocks which are proportional to a common lower body G-block are multiplied by the reduction factor and added into that common lower body residue. For a detailed diagrammatic discussion of these residue block transfers, we refer to our earlier paper \cite{Sen2015UnitaryApplications}.
\par
 The relaxed second order energy is obtained by diagonalizing the second order (H$_{eff}$)$_{\mu \nu}$ which is constructed using the converged cluster amplitudes.
 \begin{equation}
     (H_{eff})^{[2]}_{\mu \nu}=\bra{\phi_\mu}\overline{H}_\nu\ket{\phi_\nu}=H_{\mu \nu}+\sum_l H_{\mu l}t^{l(1)}_\mu \Gamma_{\mu \nu}
 \end{equation}
 where, $\Gamma$ is the Graphical Unitary Group Adapted (GUGA) transition density matrix element correlating the CSFs $\phi_\mu$ and $\phi_\nu$.
 There is also a possibility to compute the unrelaxed second order energy which circumvents the diagonalization step and it is obtained by taking an expectation value involving the unrelaxed coefficients $\{c_\mu^0\}$ only:
 \begin{equation}
     E_{unrelaxed}^{[2]}=\sum_{\mu \nu} c_\mu^{[0]}\tilde{H}_{\mu \nu}^{[2]}c_\nu^{[0]}
 \end{equation}
 The comparison between these two energies have been extensively studied by Sen et al \cite{Sen2015UnitaryApplications} and the general conclusion is that the absolute value of the unrelaxed energy is higher than the relaxed value by the order of a few millihartrees. But the PEC features and the parallelity with experimental or CI curves are reproduced well in both the approaches. This observation was corroborated by all the computations done in this paper too. We would present only the relaxed energies in the ensuing results section.
 \subsection{Projection scheme (Scheme P) to solve for cluster amplitudes:}
 When Eq. \ref{UGASSMRPTopeqn} is projected onto the set of virtual functions $\{\chi_l^\mu\}$ which is generated by singles and doubles excitation upon the CAS functions $\{\phi_\mu\}$, we get,
 \begin{equation} \label{UGASSMRPTeqn}
     \bra{\chi^l_\mu}H\ket{\phi_\mu}c^0_\mu+\bra{\chi^l_\mu}\{\contraction{}{H_0}{}{T^{(1)}_\mu}H_0T^{(1)}_\mu\}\ket{\phi_\mu}c^0_\mu-\sum_\nu\bra{\chi^l_\mu}\{\contraction{}{T^{(1)}_\mu}{}{{H_0}_{\nu\mu}}T^{(1)}_\mu {H_0}_{\nu\mu}\}\ket{\phi_\mu}c^0_\mu \\
+\sum_\nu\bra{\chi^l_\mu}\{T^{(1)}_\nu-T^{(1)}_\mu\}\ket{\phi_\mu}H_{\mu\nu}c^0_\nu=0
\end{equation}
 The solution of cluster amplitudes using equation \eqref{UGASSMRPTeqn} requires the extraction of the linearly \linebreak independent virtual functions manifold, in order to avoid singularity in the solution of the \linebreak linear simultaneous equations. The higher body operators which are obviously proportional to a particular lower body T is excluded from this extraction procedure at the very beginning itself by inspection, e.g., direct spectator active orbital scattered operators like $E_{iu}^{au}$ or doubly occupied active orbitals in exchange spectator mode like $E_{i\ u_d}^{u_d a}$. The redundancy in $\chi^l_\mu$ space has its roots in the fact that the action of multiple cluster operators on a given $\phi_\mu$ could possibly give rise to the same virtual function due to the spin-degeneracy of the $\chi_l$, e.g., the action of $\{E_i^a\}$ and $\{E_{i\ u_s}^{u_s a}\}$ (where u$_s$ refers to a singly occupied active orbital) on a model function $\ket{\cdots ji}u$, would lead to the same excited function $\ket{a\cdots j}u$. The aforementioned virtual functions could have been distinguishable in a spin-orbital based CC method as the action of $\{E_{iu}^{ua}\}$ would result in a spin flip of the electron in active orbital u resulting in a triplet excitation from i to a, but there is obviously no way to distinguish them in a spatial orbital basis. We note here that these T$_2$ operators containing 'spectator' scatterings of active orbitals are essentially equivalent to one body cluster operators when they act upon any CSF, and their inclusion is necessary to cover the complete spin-space of the virtual manifold under a given operator truncation scheme in a 'spin-free' formulation. Thus, these 'pseudo' two body cluster operators along with the T$_1$'s contribute to orbital relaxation in the perturbed wavefunction via Thouless' theorem\cite{Thouless1960StabilityTheory}. The rest of the T$_2$'s contribute to the dynamical electron correlation energy of the system.
 \par
Thus it is clear that there exists a redundancy in the virtual manifold of equation \eqref{UGASSMRPTeqn} arising from the fact that two different operators acting on a particular CSF gives the same virtual function. To get a non-singular metric in the set of projection equations, we form the overlap matrix consisting of all such operators for a fixed CSF and try to extract from it the linearly independent combinations of such operators using a singular-value decomposition (SVD).
As cluster operators which change the occupancy of different inactive orbitals can never lead to the same virtual function, we define classes of primitive operators categorized according to the number and type of inactive indices involved in the excitation and go for a SVD classwise for each CSF. Since the overlap matrix formed in this case is a Hermitian one, SVD is equivalent to orthogonal transformation of the overlap matrix and we will only be using the eigenvectors of such an \linebreak orthogonalization procedure which will be the combining coefficients($X_{Pl}$) for the orthogonal combination of the primitive operators. The number of non-zero diagonal elements left after such orthogonalization would be the number of such orthogonal combinations($l$) which again  will be equal to the number of equations at hand, thus removing redundancy in the virtual space. The scheme of extracting the linearly independent/orthogonal combinations is as follows.\\
A typical element of the overlap matrix (S$_{PQ}$) would look like:
\begin{equation}
    S_{PQ}=\bra{\phi_\mu}\{{E_P^\dagger}\}\{E_Q\}\ket{\phi_\mu}
\end{equation}
Note here, $P$ and $Q$ belong to the same primitive class. %\color{red}
$S_{PQ}$ is expressed in terms of spin-free reduced density matrices for a given CSF.\\
The SVD transformation in matrix notation is given by:
\begin{equation}
    X_{lP}^\dagger S_{PQ} X_{Ql}=(S_{ll})_d
\end{equation} %{\color{red} Why not $S_{PQ}$ ? Should be a combination of primitives.}
where the $X_{Pl}$'s are the orthogonal combining coefficients we are interested in and the primitive operators($E$) are related to the orthogonal operators($\epsilon$) as:
\begin{equation}
    \{\epsilon_l\}=\sum_P\{E_P\}X_{Pl}
\end{equation}
with
\begin{equation}
     \ket{\chi^l}=\{\epsilon^l\}\ket{\phi_\mu}
\end{equation}
Projecting onto the linearly independent manifold $\bra{\chi_l}$, the equation \eqref{UGASSMRPTeqn} takes the form: 
\begin{align}\label{proj}
\sum_{PR}X_{lP}^\dagger S_{PR}G_R=0\ \ \ \ \ \ \ \ \ \ \ \ \forall \  l \end{align}
where the $G_R$ is the primitive residue used in the amplitude equation of \eqref{UGASSMRPTopeqn}. Equation \eqref{proj} will be our projection equation.
\par
These two inequivalent approaches to solve for the excitation amplitudes is independent of the choice of H$_0$ and is also applicable to any order of perturbation right up to the full non-perturbative UGA-SSMRCC theory.
\subsection{Choice of H$^{(0)}$:}
The choice of H$^{(0)}$, i.e., the unperturbed Hamiltonian has to be such that it does not connect the model space and the virtual manifold. One might imagine that a generalized Fock operator in the natural orbital basis and defined with respect to the unperturbed contracted function $\Psi_0$ could be an acceptable choice. However this is not suitable for a perturbation theory involving an uncontracted treatment of the CAS functions $\{\phi_\mu\}$.
A much better description of H$^{(0)}$ would be multi-partitioning \cite{Zaitsevskii1995Multi-partitioningTheory} it, where we define the H$^{(0)}$ to be dependent on the particular CSF it acts on. This choice would include the effective one-body Fock potential of the doubly and singly occupied active orbitals in a given $\phi_\mu$. Since the occupancy of active orbitals differs in each reference function $\phi_\mu$, we will define the unperturbed Hamiltonian H$^{(0)}$ as,
 $$\Tilde{f}_{\mu p}^{\ q}=f_{c p}^{\ q}+\sum_{u_d\in \mu}(2V_{pu_d}^{qu_d}-V_{pu_d}^{u_dq})+\sum_{u_s\in \mu}V_{pu_s}^{qu_s}$$
 where, $f_c$ denotes the core fock operator, which is common for all $\phi_\mu$s; p,q are general orbital \linebreak indices but should belong to the same "class",i.e., inactive holes (to be denoted by i,j,..),inactive particles (to be denoted by a,b,..) or active orbitals (to be denoted by u,v,..); $u_s$ and $u_d$ correspond to singly occupied and doubly occupied orbital indices respectively.
 \par
 This effective one-body reference dependent Fock operator as H$^{(0)}$ in the UGA-SSMRPT2 theory was successfully applied by Sen et al and for a comprehensive discussion on this choice, we refer the readers to our earlier papers \cite{Sen2015UnitaryApplications,Sen2015AspectsPrototype}.\\
 %{\color{red} Are we not reporting unrelaxed numbers ? If we are, we need to write the expression for it.}
\section{Results and discussion}

%\subsection{Outline:}
We divide our applications into three categories. In the ensuing subsections we study (a) the moderate to strong avoided crossings for manifolds of LiH, Be$_2$, and the asymmetric H$_2$S$^+$ ; (b) the interlacing behaviour in the different space-spin symmetry potential energy manifolds of BC ; and (c) very weakly avoided crossing in the PECs of BeH$^{2+}$, BeF$^{2+}$, LiF and BN. %{\color{red}
\par
As emphasized in the earlier sections, each state belonging to a particular space-spin symmetry manifold have been computed separately, since in a state-specific MR theory each root is the target solution of its own specific effective Hamiltonian. There is no obvious reason for the various gradient changes and interlacing between these same symmetry states to correspond with each other. However if the formalism is rich enough to faithfully approximate each specific root it is reasonable to infer that each PEC would sense interlacing and avoided crossing to an extent demanded by the accuracy of the state-specific method. This is why a study of the behaviour of manifolds of states of a given symmetry belonging to the categories (a) to (c) is of paramount importance.
\par
In the ensuing subsections we will show that both the versions of UGA-SSMRPT2 (Scheme A and P) have faithfully described the interlacing and avoided crossing features in the PECs for all the systems studied except in LiF and BN, where we will demonstrate the failure to describe a weak ionic-covalent avoided crossing by a state-specific perturbative theory and also try to provide a rationale behind this observation. We will also present a comparative study of the two possible avenues to solve for cluster amplitudes and validate the efficacy of the sufficiency conditions we had imposed in our earlier works\cite{Sen2015UnitaryApplications,Sen2015AspectsPrototype}. GAMESS-US \cite{Barca2020RecentSystem} version 14 has been used to generate the CASSCF one and two particle MO integrals as well as the GUGA one particle transition density matrix. MOLPRO-19 \cite{MOLPRO-WIREs} has been used to generate the IC-MRCISD+Q energies in all the systems except BC whose data was kindly provided by A. Mavridis and D. Tzeli  \cite{Tzeli2001First-principlesAlC-,Tzeli2001Accurate2}. The UGA-SSMRPT2 code developed by Sen et al \cite{Sen2015UnitaryApplications} for the Scheme A was considerably modified via the exclusive usage of BLAS wherever applicable, in order to utilize its intrinsic openMP thread parallelization. In addition to that, the entire Scheme P was developed and implemented as discussed in this paper for the first time.\\
\\
\textbf{Validation of our sufficiency condition:}
\\
As discussed earlier, the rigorous method to solve for the cluster amplitudes would be via projecting the residue equations onto orthogonal virtual CSFs (Scheme P), and this could be bypassed if we choose a sufficiency condition such that all individual amplitudes of the excitation part of H$_{eff}$ are set to zero (Scheme A). Scheme A has been extensively used by Sen et al in his applications \cite{Sen2015AspectsPrototype} and has been shown to be very promising. In what follows, we validate the relative performance of the computationally cheaper Scheme A against the more involved Scheme P using MRCISD+Q as the benchmark method.
\begin{table}[H]
\centering
    \begin{tabular}{|c|c|c|c|} 
     \hline
    Molecule, & UGA-SSMRPT2 Scheme & NPE wrt MRCI+Q & MAD wrt MRCI+Q \\%& D$_e$ \\
    State and Basis&&(in a.u.)&(in a.u.)\\%&(in eV)\\
    \hline
    LiH 1$^1\Sigma^+$&A&4.87$\times10^{-4}$&1.60$\times10^{-4}$\\%&\\
    aug-cc-pVTZ&P&4.87$\times10^{-4}$&1.60$\times10^{-4}$\\%&\\
    \hline
    LiH 2$^1\Sigma^+$&A&4.58$\times10^{-4}$&1.52$\times10^{-4}$\\%&\\
    aug-cc-pVTZ&P&4.58$\times10^{-4}$&1.52$\times10^{-4}$\\%&\\
    \hline
     Be$_2$ 1$^1\Sigma_g^+$&A&2.44$\times10^{-3}$&2.85$\times10^{-4}$\\%&\\
    aug-cc-pVQZ&P&2.45$\times10^{-3}$&2.95$\times10^{-4}$\\%&\\
    \hline
    BeH$^{2+}$ 1$^2\Sigma^+$&A&5.60$\times10^{-5}$&2.13$\times10^{-5}$\\%&\\
    aug-cc-pVDZ&P&5.60$\times10^{-5}$&2.13$\times10^{-5}$\\%&\\
    \hline
    BeH$^{2+}$ 2$^2\Sigma^+$&A&4.47$\times10^{-4}$&2.26$\times10^{-5}$\\%&\\
    aug-cc-pVDZ&P&4.47$\times10^{-4}$&2.26$\times10^{-5}$\\%&\\
    \hline
    BeH$^{2+}$ 3$^2\Sigma^+$&A&4.46$\times10^{-4}$&2.05$\times10^{-5}$\\%&\\
    aug-cc-pVDZ&P&4.46$\times10^{-4}$&2.05$\times10^{-5}$\\%&\\
    \hline
     BC 1$^4\Sigma^-$&A&7.84$\times10^{-3}$&2.19$\times10^{-3}$\\%&\\
    cc-pV5Z(-h)&P&7.63$\times10^{-3}$&2.08$\times10^{-3}$\\%&\\
    \hline
    \end{tabular}
\caption{Relative performance of the two inequivalent schemes A and P in UGA-SSMRPT2}%[experimental values of D$_e$ given in parentheses]}
\label{A vs P}
\end{table}
We find that these two schemes have a very similar performance with respect to the computed potential energy curves.
\par
 For the sake of clarity in the figures and to avoid cluttering of data points in them, hereafter we will present the UGA-SSMRPT2 Scheme A numbers in the PECs to demonstrate the proper avoided crossing and interlacing behaviour predicted by our theory.\\
 \\
\textbf{Validation of the PEC features:}
\\
 In this subsection, we compare the avoided crossing features given by our theory with the MRCISD+Q numbers. The dissociation energy computed by us is compared with experimental data wherever available.
\begin{table}[H]
\centering
    \begin{tabular}{|c|c|c|c|} 
     \hline
    Molecule & States & UGA-SSMRPT2 & MRCI+Q \\
    &&(in a.u.)&(in a.u.)\\
    \hline
    BeF$^{2+}$&1$^2\Pi$,2$^2\Pi$&22.23&20.55\\
    \hline
    BeH$^{2+}$&2$^2\Sigma^+$,3$^2\Sigma^+$&52.05&51.80\\
    \hline
    BC&2$^2\Pi$,3$^2\Pi$&2.75&2.65\\
    \hline
     \end{tabular}
\caption{Comparison of the avoided crossing region for MRCI+Q and UGA-SSMRPT2}
\label{Avoided crossing comparison}
\end{table}
\begin{table}[H]
\centering
    \begin{tabular}{|c|c|c|c|} 
     \hline
    Molecule & States & UGA-SSMRPT2 & MRCI+Q \\
    &&(in eV)&(in eV)\\
    \hline
    LiH&1$^1\Sigma^+$,2$^1\Sigma^+$&1.23&1.19\\
    \hline
    BeH$^{2+}$&2$^2\Sigma^+$,3$^2\Sigma^+$&1.60$\times$10$^{-4}$&1.99$\times$10$^{-4}$\\
    \hline
    BC&2$^2\Pi$,3$^2\Pi$&0.19&0.14\\
    \hline
     \end{tabular}
\caption{Comparison of the avoided crossing energy gap for MRCI+Q and UGA-SSMRPT2}
\label{Avoided crossing gap}
\end{table}
\begin{table}[H]
\centering
    \begin{tabular}{|c|c|c|c|c|} 
     \hline
    Molecule & States & UGA-SSMRPT2 & MRCI+Q & Expt. \\
    &&(in eV)&(in eV)&(in eV)\\
    \hline
    LiH&1$^1\Sigma^+$&2.522&2.522&2.515\cite{Partridge1981TheoreticalLiHb}\\
    \hline
    LiH&2$^1\Sigma^+$&1.075&1.079&1.076\cite{Partridge1981TheoreticalLiHb}\\
    \hline
    Be$_2$&1$^1\Sigma_g^+$&0.138&0.129&0.116\cite{Schmidt2010ElectronicBe2}\\
    \hline
    BeF$^{2+}$&1$^2\Pi$&1.93&2.03&1.92\cite{Kolbuszewski1993PredictingBeF2+}\\
    \hline
    BC&1$^4\Sigma^-$&4.554&4.375&4.597\cite{Tzeli2001First-principlesAlC-}\\
    \hline
     \end{tabular}
\caption{Dissociation energy for the computed states with UGA-SSMRPT2}
\label{Dissociation energy}
\end{table}
We see that the interlacing and avoided crossing features are reasonably well reproduced by our state-specific theory. In the following subsections, we show the absolute PEC of each system studied by our theory and demonstrate the aforementioned PEC features. \par We also find that our computed dissociation energies tally very well with the corresponding experimental values and this further validates the rigorous size-extensivity present in our formulation. This gives us a glimpse that the UGA-SSMRPT2 is a good candidate for gaining insight to dissociation and bonding phenomena without sacrificing too much numerical accuracy. 
\par
Keeping these observations in mind, we proceed to describe the molecular states studied by us in greater detail.
\subsection{Molecular states exhibiting strong to moderately strong avoided crossings:}
\noindent
\textbf{\underline{LiH}}\\
\\
LiH bond dissociation has always been the first test case that comes to mind for most electronic structure theories. The main reasons are: (i) it shows a reasonable multireference character as soon as it is stretched away from equilibrium and it is in full force as we approach the fragmentation limit (ii) it is easy to have a FCI comparison with reasonably large bases (iii) the 1st excited state shows reasonably strong avoided crossing with the ground state along with considerable variation in dipole moment.
\par
To study the moderately strong avoided crossing between the ground and 1st excited $^1\Sigma^+$ states, a (2,2) CAS containing the CSFs \{$\sigma^2,\sigma \sigma^*\ \text{and}\ \sigma^{*2} $\} is sufficient. The ground and 1st excited states are dominated by the ionic and covalent configurations respectively before this avoided crossing, and the configurations swap after that point.
\par
However, the correct fragmentation limit of the 1st excited $^1\Sigma^+$ state is not ionic but a covalent H ($^2$S) and Li ($^2$P) channel. This involves passing through a second avoided crossing and this phenomenon can be described well using an appropriately large basis \cite{Partridge1981TheoreticalLiHb} alongwith the inclusion of 2p and 3d orbitals of Li in the CAS, essentially necessitating a (2,10) CAS \cite{Theis2010MolecularTheory}. All orbitals were correlated in our computation.%{\color{red} Which CAS is this ? Also mention CAS in all captions.}
\begin{figure}[H]
     \centering
        \includegraphics[width=\textwidth]{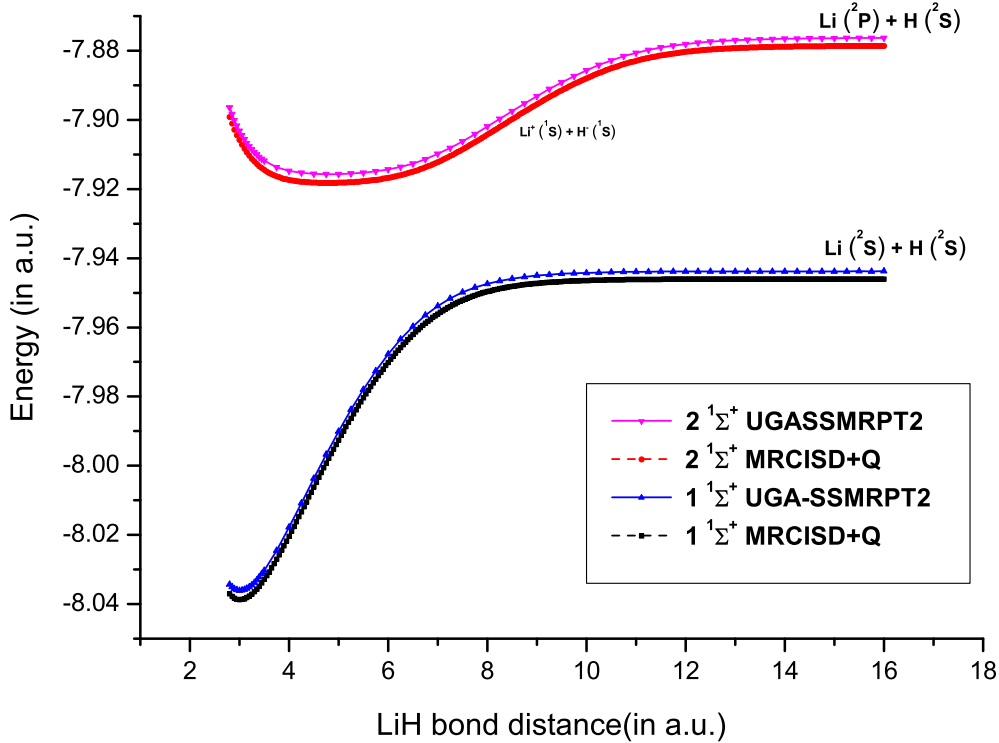}
        \caption{Ground and 1st excited $\ ^1\Sigma^+$  states of LiH computed using the aug-cc-pVTZ basis and a (2,10) CAS}
        \label{LiH_PES_acct}
\end{figure}
From Fig. \ref{LiH_PES_acct}, we find that the gradient changes in our UGA-SSMRPT2 curve follow very closely with the corresponding MRCISD+Q curve. Here we again stress the fact that for MRCISD+Q we get both states as the roots of the same Hamiltonian, while in our theory the two states are solutions of two different effective Hamiltonians. The non-parallelity error with respect to MRCISD+Q is of the order of 10$^{-4}$ hartrees as seen in Table \ref{A vs P} and the gap between the two states taken at 7.50 bohrs is very much comparable between the two methods as seen in Table \ref{Avoided crossing gap}. From Table \ref{Dissociation energy}, we note that the dissociation energies computed by us differs from the experimental values \cite{Partridge1981TheoreticalLiHb} by a few milli-eV.\\
\\
\textbf{\underline{Be$_2$}}\\
\\
Be$_2$ has a very unusual bonding which is hard to take account of, at a non-correlated level. It has been the topic of research for many decades both theoretically and experimentally. Although the bond is weaker than an average chemical bond, it is far stronger than one caused by van der Waal dispersion forces. Be$_2$ has two distinct features in its dissociation curve: (i) a deep minimum at short distance which can be described theoretically only after the inclusion of dynamical electron correlation (ii) a significant gradient change, owing to a very strong avoided crossing with an energetically much higher non-bonding van der Waal state, before fragmentation into two Be($^1$S) atoms.
\par
The unexpected bonding minimum in Be$_2$ is attributed to the promotion of an s electron to p orbital, resulting in a Be($^3$P) state\cite{Lepett1990InteractionPromotion}. The avoided crossing between the non-bonding state consisting of two 1s$^2$ 2s$^2$ Be atoms and the bonding state consisting of two 1s$^2$ 2s$^1$ 2p$^1$ Be atoms results in the deep bonding minimum of the ground state around 4.50 bohrs.%{\color{red} Not demonstrated in the plots}. 
There is also a strong angular correlation from the 3d$_\pi$ orbitals to this sp hybrid state. Thus, the binding in Be$_2$ can be attributed to the dynamical correlation effect, which is very rare in case of similar dimer systems.  %coupled with the strong near degeneracy effect of the s$^2$ to p$^2$ {\color{red} Not $s^1p^1$ ?} excitation. 
This is also why both the RHF and the CASSCF PEC of Be$_2$ show no bonding minimum.
\par
The change in slope of the ground state PEC around 6.0 a.u. is ascribed to the contribution of 3d and additionally the presence of f and g orbitals. This is exemplified by the basis set dependence of the ground state PEC which has been studied previously by Ruedenberg et al \cite{Schmidt2010ElectronicBe2} and Kalemos \cite{Kalemos2016TheBe3}. It has been shown that the inclusion of upto p functions in the basis does not give the deep minimum at all in the CI curves. The inclusion of upto d functions gives not only the bonding well but also a second shallow minimum as an artefact. The experimental PEC is reproduced only after including the diffuse f and g functions in the basis for Be.
\par
We reproduce these features %{\color{red}
indicating the basis set effects with our perturbation theory using a minimal CAS (4,4) and delocalized canonical CASSCF orbitals. We employ both the aug-cc-pVDZ (containing upto d functions) and the aug-cc-pVQZ (containing upto g functions) bases. All orbitals were correlated in our computation. %However, to achieve a size consistent fragmentation limit, a (8,8) CAS is necessary.
\begin{figure}[H]
     \centering
        \includegraphics[width=\textwidth]{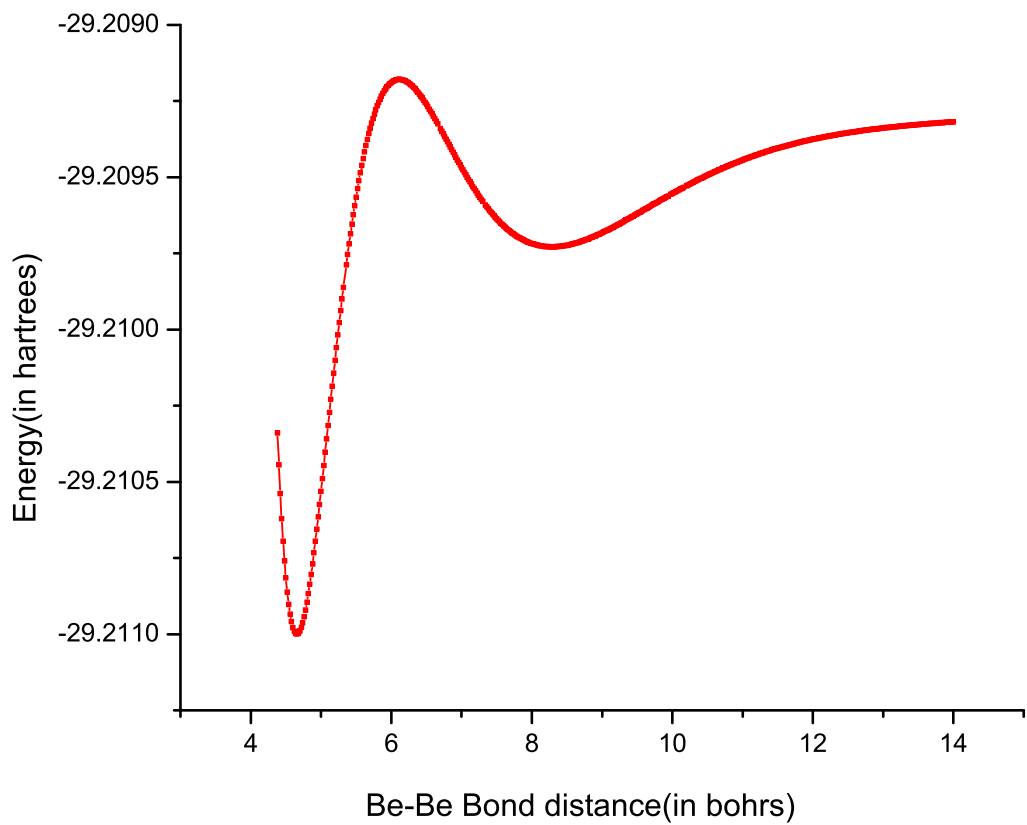}
        \caption{Ground $\ ^1\Sigma_g^+$  state of Be$_2$ computed by UGA-SSMRPT2 using the aug-cc-pVDZ basis and a (4,4) CAS}
        \label{Be2_PES_accd}
\end{figure}
\begin{figure}[H]
     \centering
        \includegraphics[width=\textwidth]{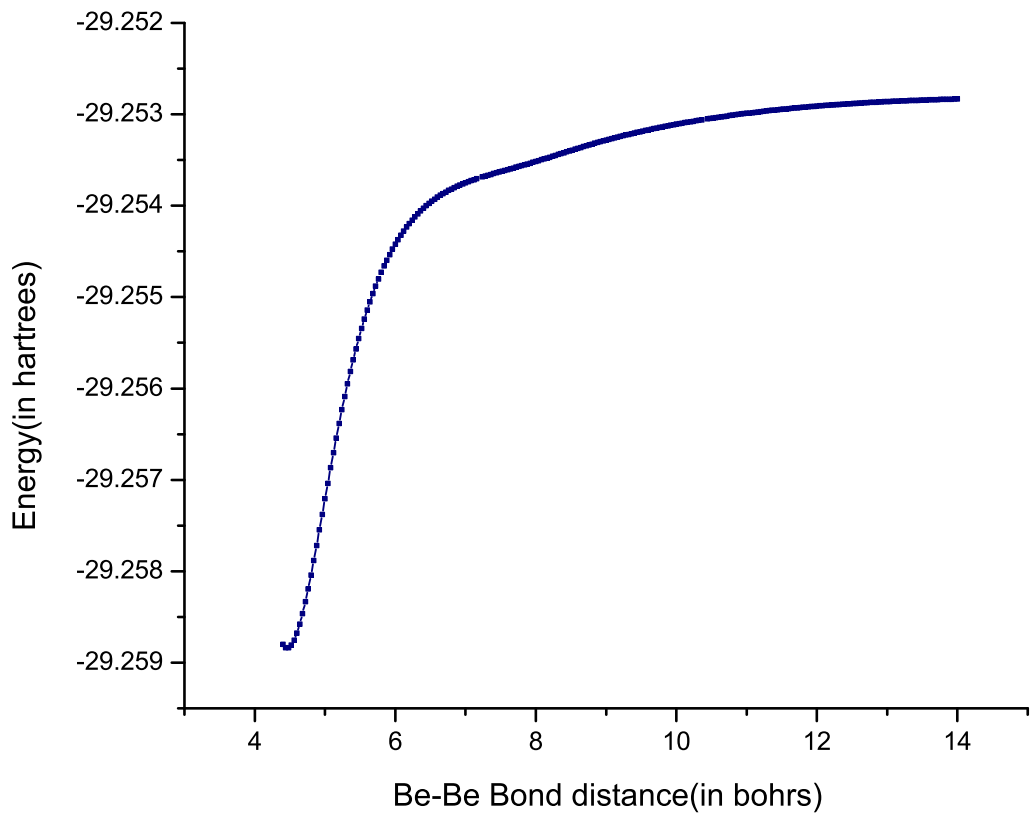}
        \caption{Ground $\ ^1\Sigma_g^+$  state of Be$_2$ computed by UGA-SSMRPT2 using the aug-cc-pVQZ basis and a (4,4) CAS. The emergence of the shallow minimum is clearly manifest}
        \label{Be2_PES_accq}
\end{figure}
\begin{figure}[H]
     \centering
        \includegraphics[width=\textwidth]{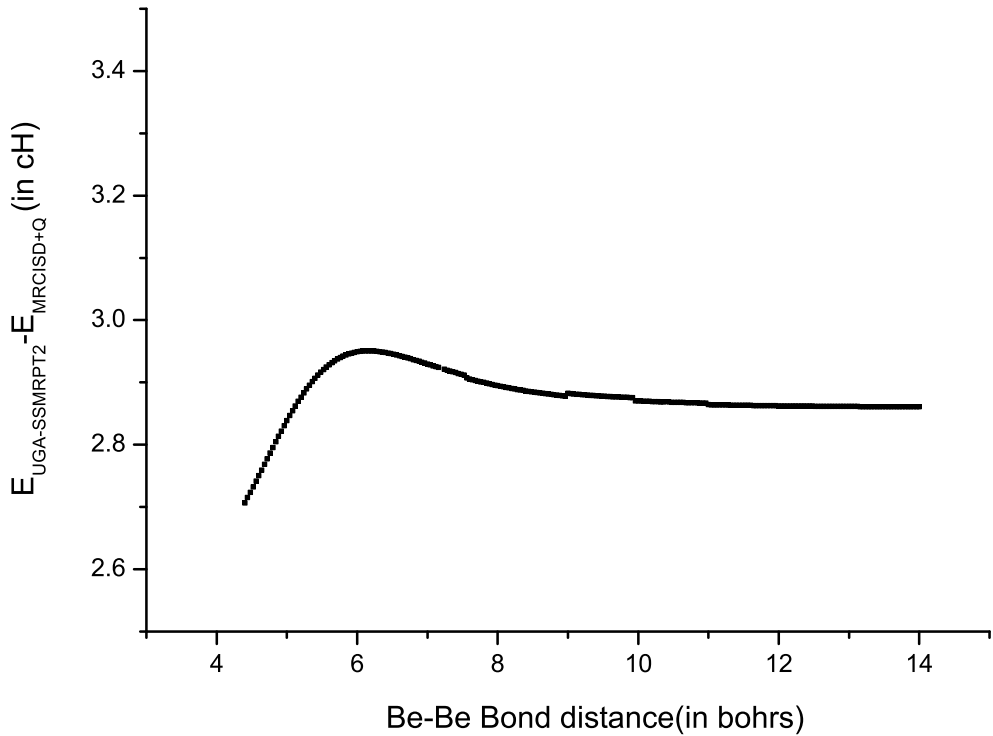}
        \caption{Energy difference for ground $\ ^1\Sigma_g^+$  state of Be$_2$ between UGA-SSMRPT2 and MRCISD+Q using the aug-cc-pVQZ basis and a (4,4) CAS}
        \label{Be2_accq_energy_diff}
\end{figure}
From the Figs. \ref{Be2_PES_accd} and \ref{Be2_PES_accq}, we see that even at a much lower order of correlation the UGA-SSMRPT2 has reproduced all the distinctive features in the PEC of Be$_2$ alongwith the pronounced basis set effects on it. The non-parallelity between our computed PEC and that by MRCISD+Q is of the order of millihartrees as seen in the above curves \linebreak and in Table \ref{A vs P}. Figure \ref{Be2_accq_energy_diff} further demonstrates this difference with respect to the Be-Be bond distance.
\par
The dissociation energy computed, using the delocalized (4,4) CASSCF orbitals, differs from the experimentally obtained value by 0.022 eV as seen in Table \ref{Dissociation energy}. As noted before, the minimal CAS was enough for our theory to simulate all the essential characteristics of the ground state PEC on which we are focusing in this paper. However, to achieve a proper size consistent fragmentation limit and possibly improve upon the value of dissociation energy, an (8,8) CAS containing all the s,p orbitals on each atom should be employed. This ensures proper localization of the MOs on each Be atom which in turn results in the size consistency error to be of the order of 10$^{-1}\mu$H. This issue of size consistency in homonuclear diatomic systems has been exclusively discussed in our earlier paper\cite{Sen2015AspectsPrototype}.\\
\\
\textbf{\underline{H$_2$S$^+$:}}\\
\\
The 1st two excited $^2$A$^\prime$ states of asymmetric H$_2$S$^+$ cation have been an interesting test case for multireference methods and has been studied by Li et al \cite{Li2005TheTheory} and Datta et al \cite{Datta2011TheTheory}  previously. It has been studied as a function of change in $\angle$H-S-H bond angle and exhibits a moderately weak avoided crossing between the two states, giving rise to two minima in the 1$^2$A$^\prime$ state and one global minimum in the 2$^2$A$^\prime$ state.
\par
In our UGA-SSMRPT2 computation, we use a (3,2) CASSCF function state-averaged over the 1st two $^2$A$^\prime$ states. We employ the Dunning cc-pVQZ basis for both H and S atoms. The bond lengths of H-S are fixed at 1.595 \AA\ and 1.399 \AA. Five core orbitals were kept frozen during the computation of dynamic correlation.
\begin{figure}[H]
     \centering
        \includegraphics[width=\textwidth]{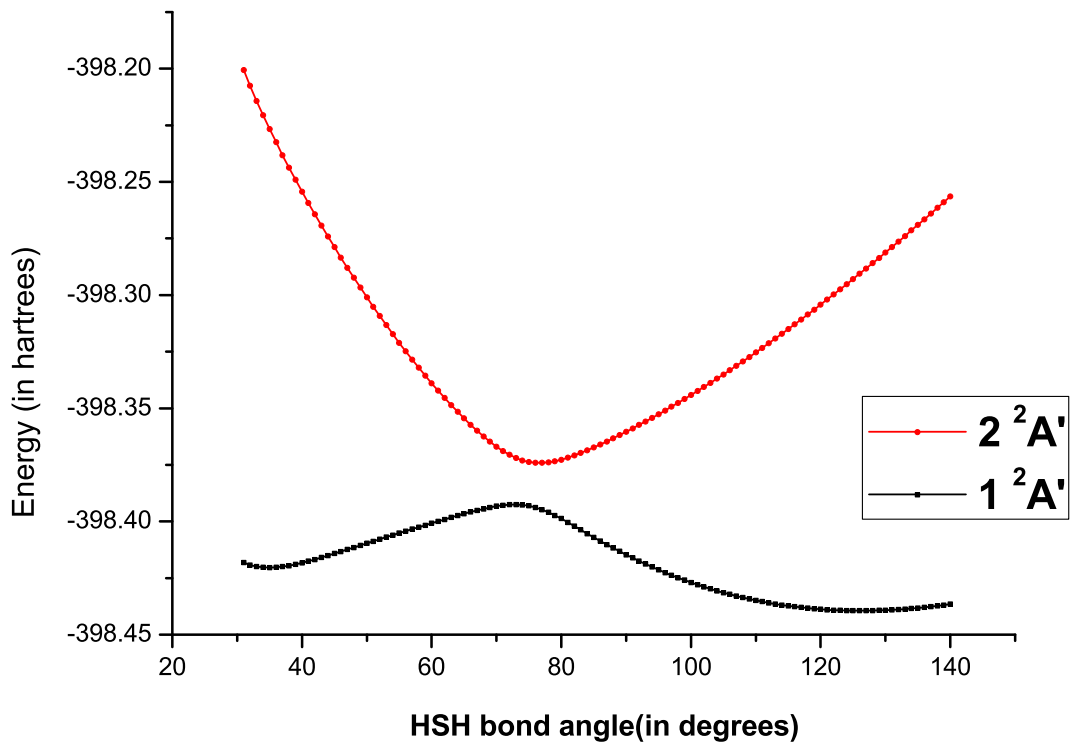}
        \caption{Ground and 1st excited $^2$A$^\prime$ states of asymmetric H$_2$S$^+$ computed by UGA-SSMRPT2 using cc-pVQZ basis and a (3,2) CAS}
        \label{H2S+_ampl}
\end{figure}
From Fig. \ref{H2S+_ampl}, we see that the change in gradient with change in the bond angle $\angle$H-S-H and the moderately weak avoided crossing between the two states is neatly reproduced by our theory.
\subsection{Interlacing within a manifold of states of same space-spin symmetry in BC:}
The potential energy profiles of the ground state $^4\Sigma^-$ symmetry manifold of BC along with it's excited states of varying space-spin symmetries provide a comprehensive test case for our theory, to be able to reproduce the various interlacing and avoided crossings in each manifold. We follow the work done by Mavridis et al \cite{Tzeli2001Accurate2,Tzeli2001First-principlesAlC-} where they have computed the MRCISD(+Q) profiles of each manifold using a high level basis set, cc-pV5Z(-h). A (7,8) CAS was employed for each manifold and the starting function was a dynamically weighted state averaged CAS function over all the states computed in an individual space-spin symmetry manifold. The $\Pi$ states were computed using a reduced point group symmetry of C$_2$ in order to preserve the xy degeneracy of the $\pi$ MOs. The two 1s-dominant molecular orbitals were kept frozen during all our computations.
\begin{figure}[H]
     \centering
        \includegraphics[width=\textwidth]{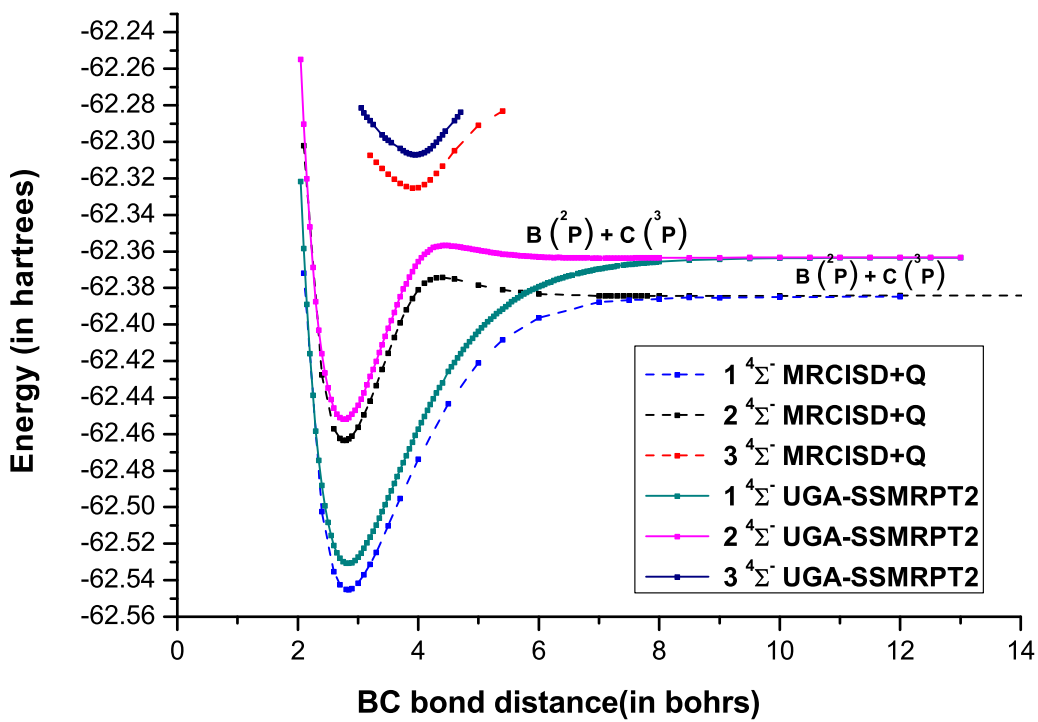}
        \caption{The $^4\Sigma^-$ manifold of BC computed by UGA-SSMRPT2 and MRCISD+Q using the cc-pV5Z(-h) basis and a (7,8) CAS}
        \label{4sigma-_manifold_BC}
\end{figure}
\begin{figure}[H]
     \centering
        \includegraphics[width=\textwidth]{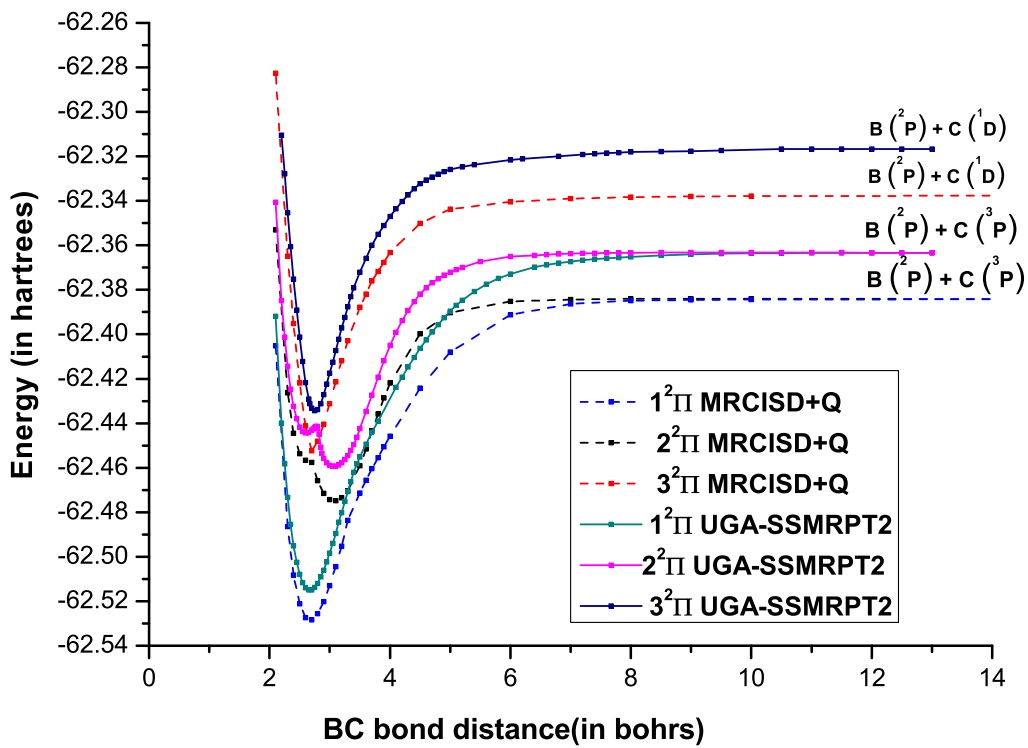}
        \caption{The $^2\Pi$ manifold of BC computed by UGA-SSMRPT2 and MRCISD+Q using the cc-pV5Z(-h) basis and a (7,8) CAS. See Figs. \ref{2pi_1_2_zoomed_BC} and \ref{2pi_2_3_zoomed_BC} for a zoomed display of the avoided crossings between 1$^2\Pi$, 2$^2\Pi$ and 2$^2\Pi$, 3$^2\Pi$ states respectively.}
        \label{2pi_manifold_BC}
\end{figure}
\begin{figure}[H]
     \centering
        \includegraphics[width=\textwidth]{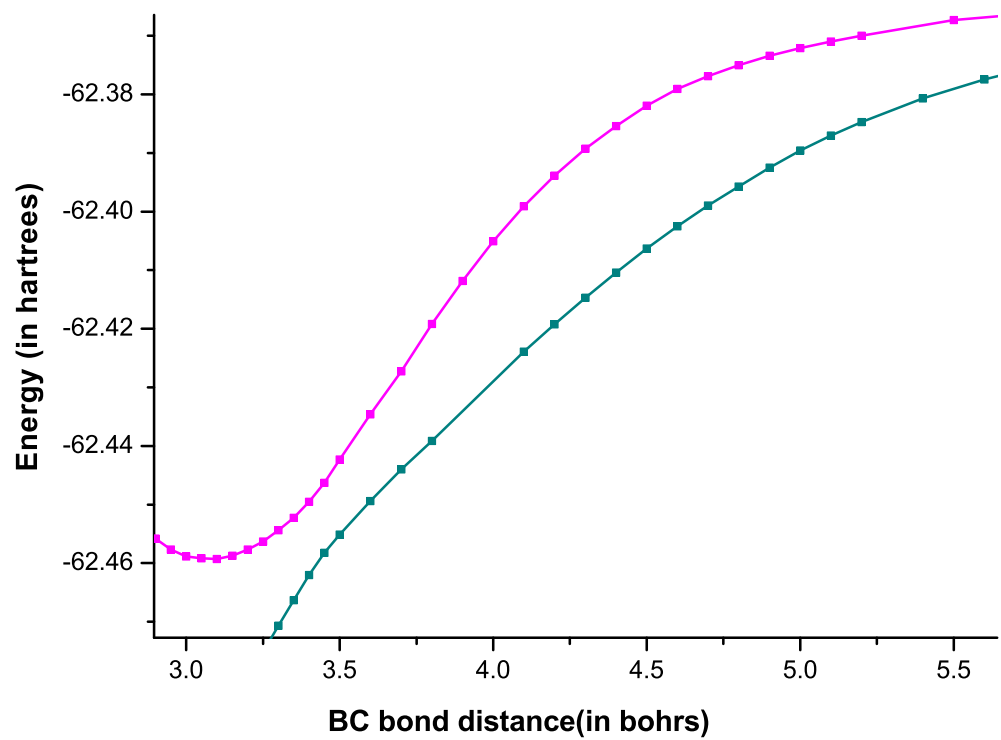}
        \caption{The avoided crossing region between the ground and 1st excited $^2\Pi$ states of BC computed by UGA-SSMRPT2 using the cc-pV5Z(-h) basis}
        \label{2pi_1_2_zoomed_BC}
\end{figure}
\begin{figure}[H]
     \centering
        \includegraphics[width=\textwidth]{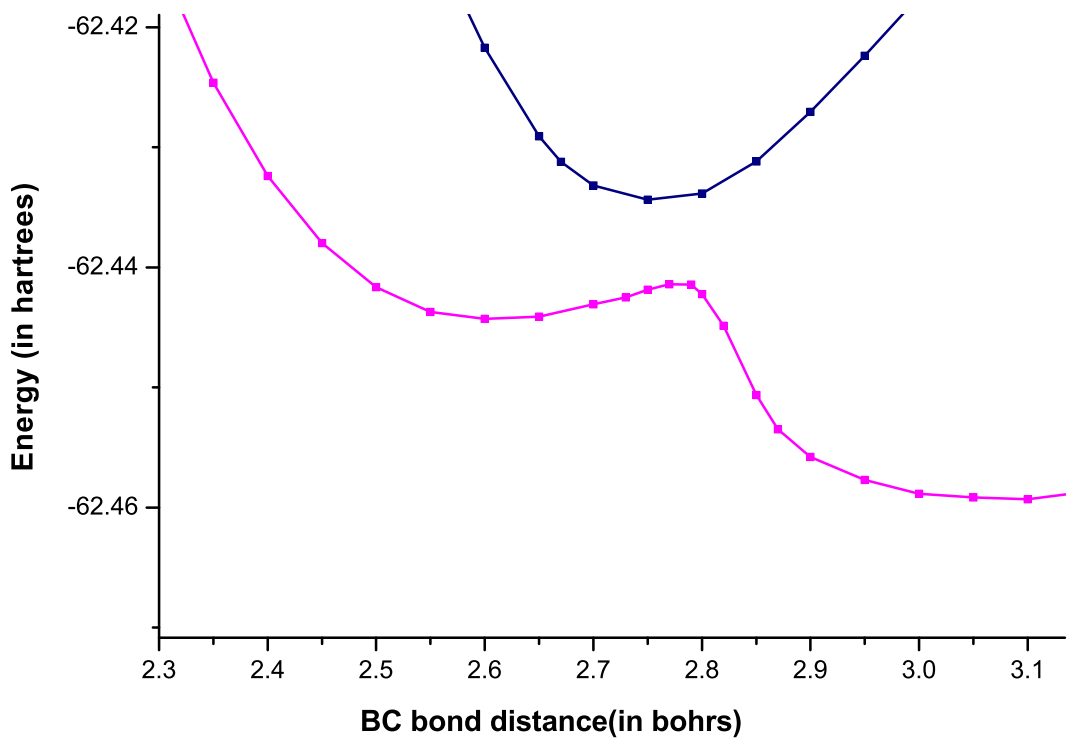}
        \caption{The avoided crossing region between the 1st and 2nd excited $^2\Pi$ states of BC computed by UGA-SSMRPT2 using the cc-pV5Z(-h) basis}
        \label{2pi_2_3_zoomed_BC}
\end{figure}
\begin{figure}[H]
     \centering
        \includegraphics[width=\textwidth]{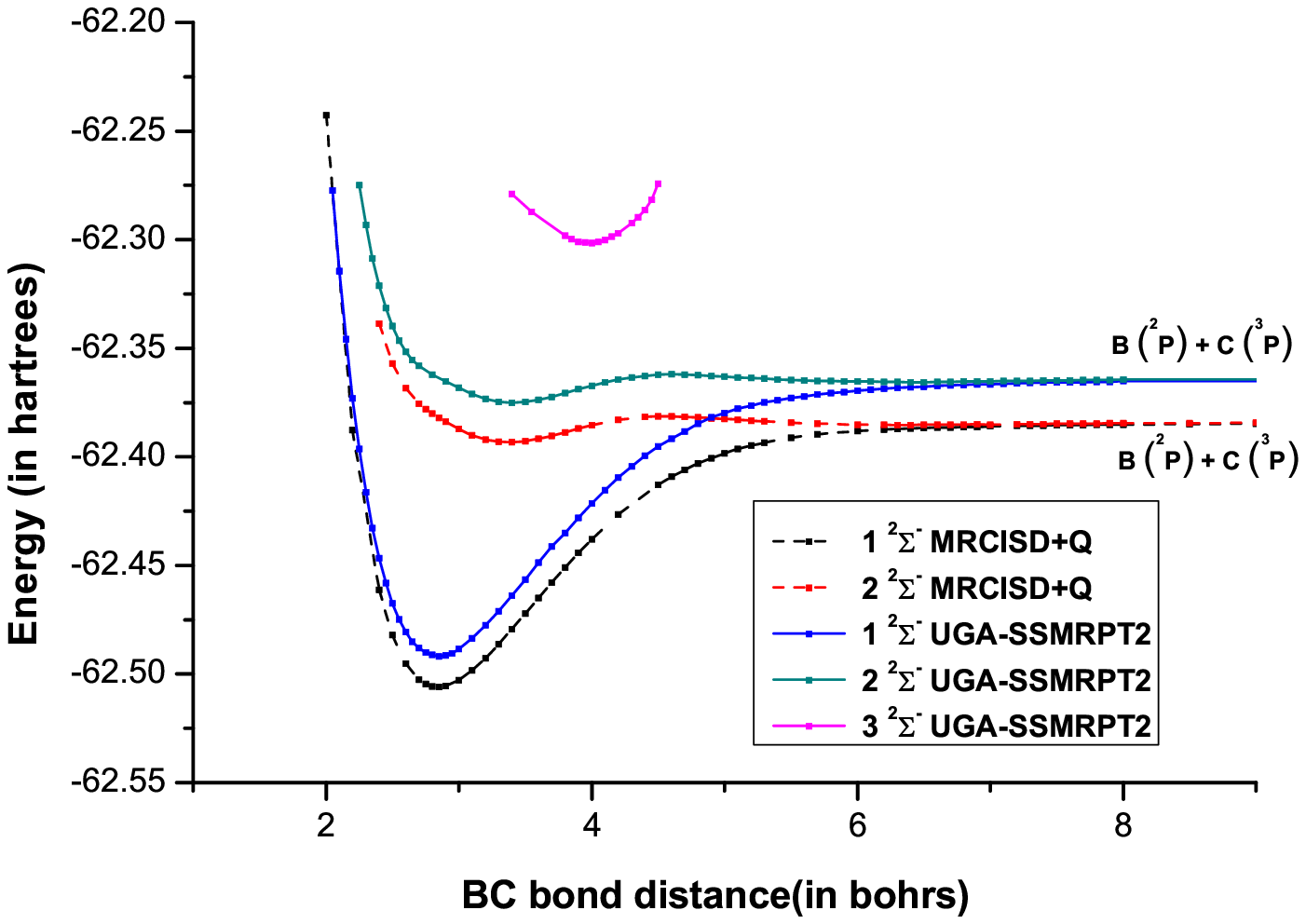}
        \caption{The $^2\Sigma^-$ manifold of BC computed by UGA-SSMRPT2 and MRCISD+Q using the cc-pV5Z(-h) basis and a (7,8) CAS}
        \label{2sigma-_manifold_BC}
\end{figure}
\begin{figure}[H]
     \centering
        \includegraphics[width=\textwidth]{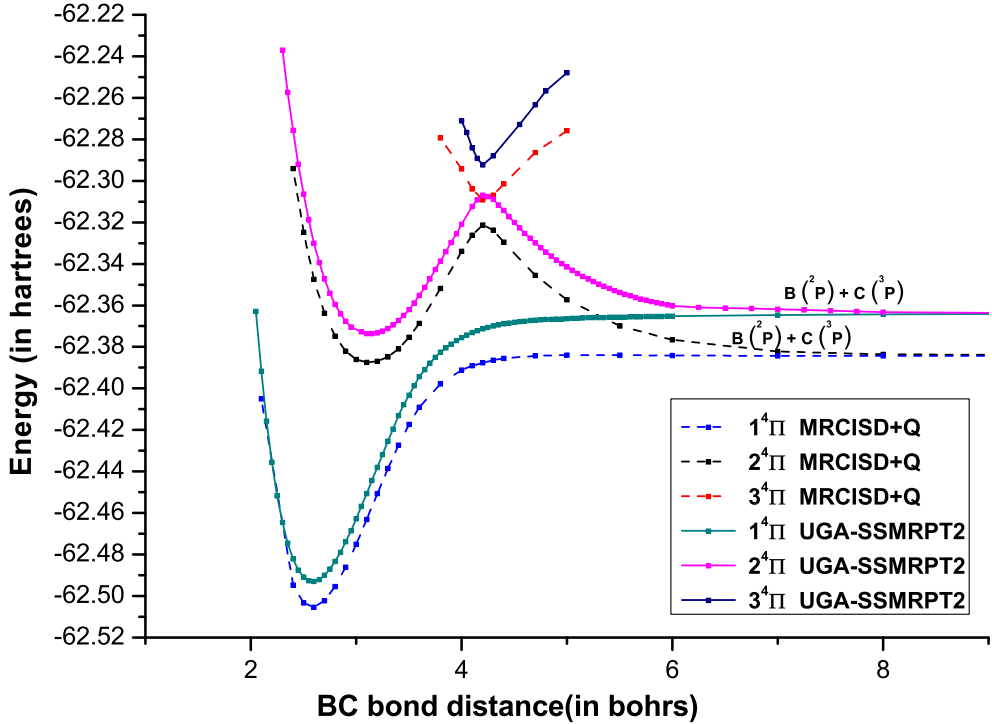}
        \caption{The $^4\Pi$ manifold of BC computed by UGA-SSMRPT2 and MRCISD+Q using the cc-pV5Z(-h) basis and a (7,8) CAS}
        \label{4pi_manifold_BC}
\end{figure}
All the above Figs. \ref{4sigma-_manifold_BC}-\ref{4pi_manifold_BC} demonstrate the accurate reproduction of the multiple interlacing and avoided crossing behaviour in each space-spin symmetry manifold of states. The computed curves differ from MRCISD+Q values computed by Mavridis et al \cite{Tzeli2001First-principlesAlC-,Tzeli2001Accurate2} by a few millihartrees in terms of non-parallelity error, as is shown in Table \ref{A vs P}.
\par
The moderately weak avoided crossing in the $^2\Pi$ manifold also occurs at a distance very close to that seen in the MRCISD+Q computations alongwith a comparable energy gap in that region, as can be seen in the Tables \ref{Avoided crossing comparison} and \ref{Avoided crossing gap}. The experimental dissociation energy was available only for the $^4\Sigma^-$ ground state of BC and the value computed by us differs from it by just ~0.04 eV as shown in Table \ref{Dissociation energy}.
\subsection{Molecular states exhibiting very weak avoided crossings:}
\noindent
\textbf{\underline{BeF$^{2+}$:}}\\
\\
BeF$^{2+}$ is one of the few diatomic dications which is thermodynamically stable \cite{Kolbuszewski1993PredictingBeF2+} in its ground $^2\Pi$ state. This system is well-suited to test our theory for weakly avoided crossings. In fact, the ground state has a low energy barrier of dissociation to Be$^+$ and F$^+$ due to a very weak avoided crossing with the 1st excited $^2\Pi$ state.
\par
We employ the Dunning cc-pVDZ basis and a (5,4) CAS to generate the ground and 1st excited $^2\Pi$ states based on a state-averaged CASSCF function. All orbitals were correlated in our computation. A reduced symmetry group of C$_2$ was used in order to conserve the x-y degeneracy of the $\pi$ orbitals. Thus our starting function had to be state-averaged over 2 pairs of degenerate states to obtain the 'symmetry-pure' state.%{\color{red} Some more discussion necessary. Quote numbers from table II and IV. You can do this for all the molecules actually.}
\begin{figure}[H]
     \centering
        \includegraphics[width=\textwidth]{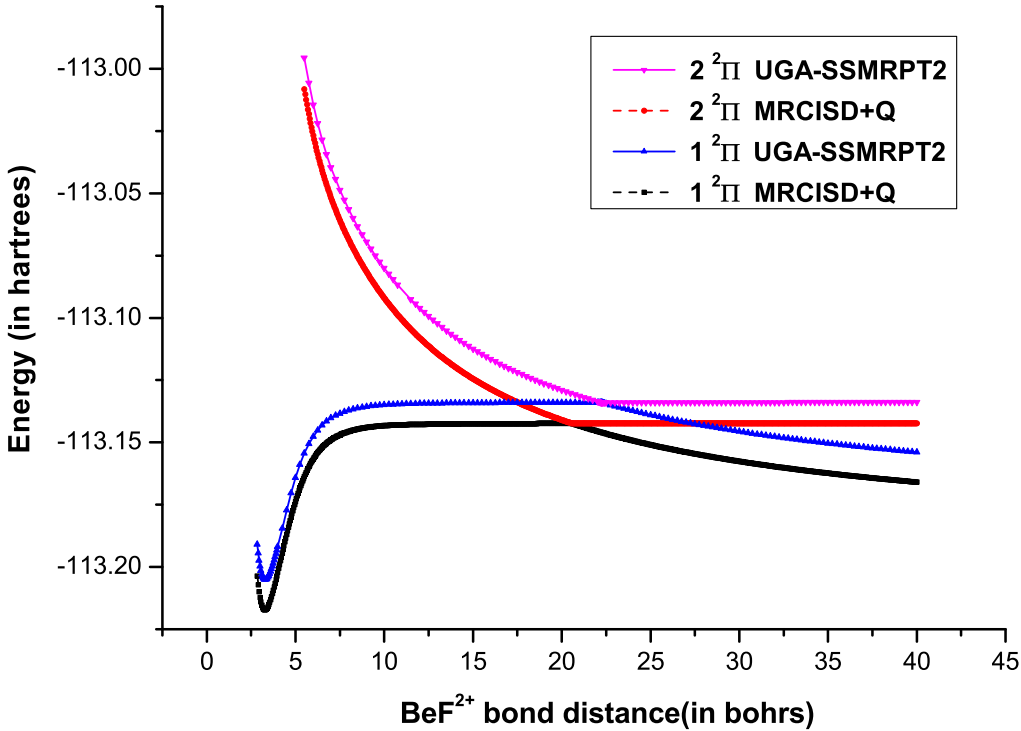}
        \caption{Ground and 1st excited $^2\Pi$  states of BeF$^{2+}$ computed using the cc-pVDZ basis and a (5,4) CAS. See Fig. \ref{BeF2+_zoom_ccd} for a zoomed display of the very weak avoided crossing.}
        \label{BeF2+_PES_ccd}
\end{figure}
\begin{figure}[H]
     \centering
        \includegraphics[width=\textwidth]{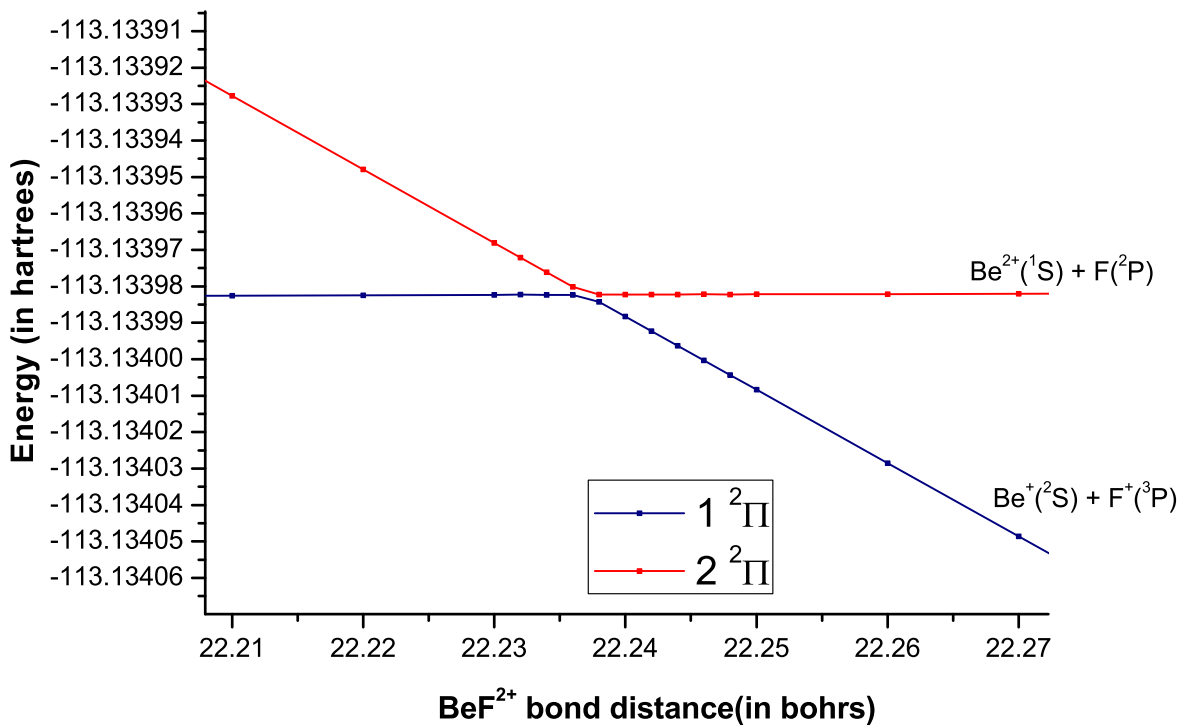}
        \caption{The avoided crossing region between the ground and 1st excited $^2\Pi$  states of BeF$^{2+}$ computed by UGA-SSMRPT2 using the cc-pVDZ basis}
        \label{BeF2+_zoom_ccd}
\end{figure}
The origin of the thermodynamic stability of BeF$^{2+}$ in its ground state is well depicted by our computations, as seen in Fig. \ref{BeF2+_PES_ccd}. The very weak avoided crossing region is shown further in a zoomed image of the PEC in Fig. \ref{BeF2+_zoom_ccd}. From Table \ref{Avoided crossing comparison}, we find that the avoided crossing between the two states computed by our theory occurs about 1.60 bohr farther than that in the MRCISD+Q curves. But the overall gradient changes in both curves are adequately described by our theory. The dissociation energy of the ground state differs by 0.01 eV from the experimental value, as seen in Table \ref{Dissociation energy}.\\
\\
\textbf{\underline{BeH$^{2+}$:}}\\
\\
BeH$^{2+}$ is a short-lived, metastable dication which has been studied both theoretically and experimentally in the last decade \cite{Farjallah2012TheoreticalFunctions}. The ground $^2\Sigma^+$ state shows a moderately strong avoided crossing with the 1st excited $^2\Sigma^+$ state at short distance, resulting in rapid dissociation to the repulsive Be$^+$($^2$S) and H$^+$ fragments. The 1st excited $^2\Sigma^+$ state seems to have stable dissociation limit, but at a very large distance ($\sim$50 a.u.) it encounters an extremely weak avoided crossing with the 2nd excited $^2\Sigma^+$ state giving rise to the repulsive Be$^+$($^2$P) and H$^+$ fragments. We study both these avoided crossings with our theory using the Dunning aug-cc-pVDZ basis and employing a (1,4) CASSCF starting function which is state-averaged over the 1st three $^2\Sigma^+$ states. All orbitals were correlated in our computation.
\begin{figure}[H]
     \centering
        \includegraphics[width=\textwidth]{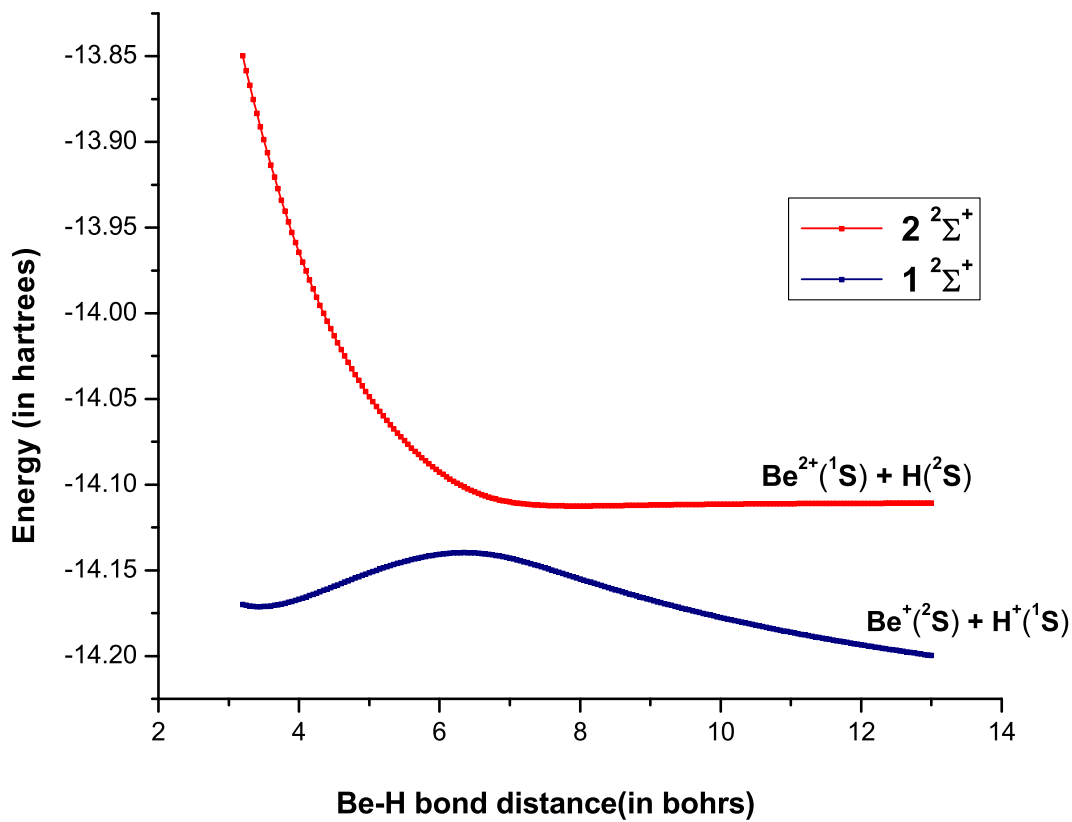}
        \caption{Ground and 1st excited $^2\Sigma^+$  states of BeH$^{2+}$ computed by UGA-SSMRPT2 using the aug-cc-pVDZ basis and a (1,4) CAS}
        \label{BeH2+_1_2}
\end{figure}
\begin{figure}[H]
     \centering
        \includegraphics[width=\textwidth]{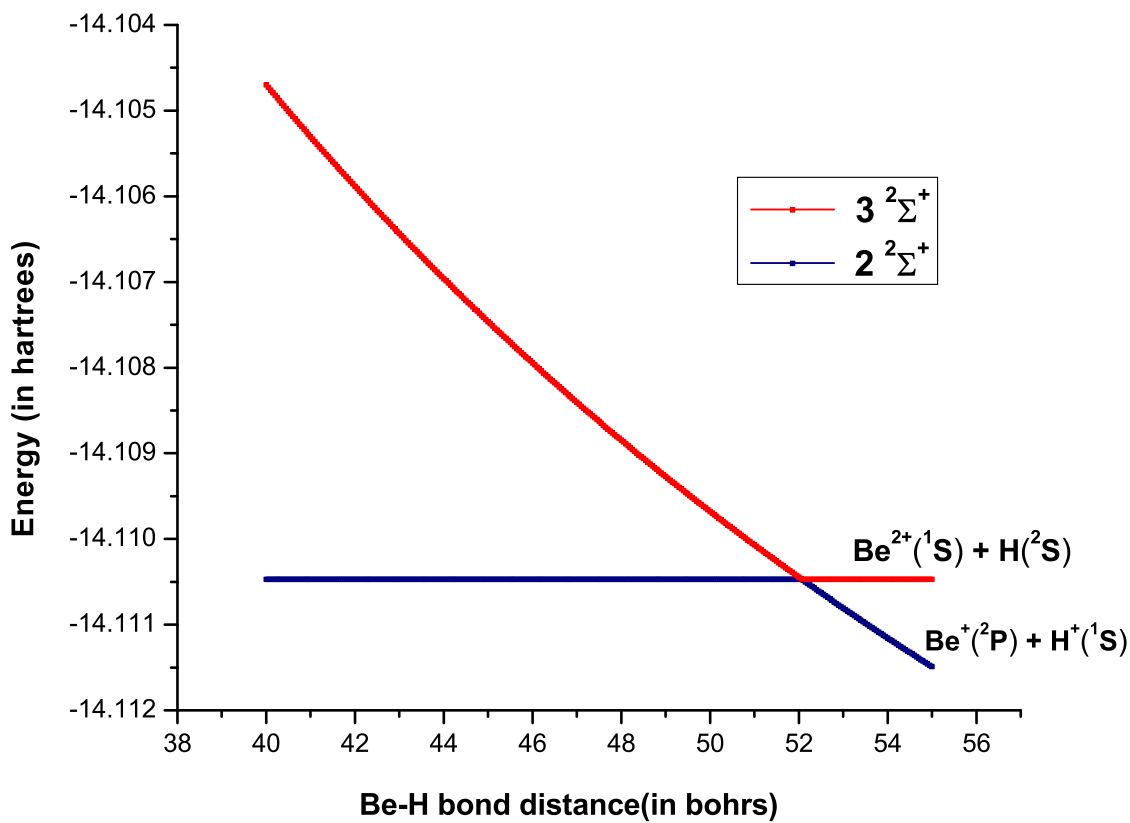}
        \caption{1st and 2nd excited $^2\Sigma^+$  states of BeH$^{2+}$ computed by UGA-SSMRPT2 using the aug-cc-pVDZ basis and a (1,4) CAS. See Fig. \ref{BeH2+_2_3_zoomed} for a zoomed display of the very weak avoided crossing region.}
        \label{BeH2+_2_3}
\end{figure}
\begin{figure}[H]
     \centering
        \includegraphics[width=\textwidth]{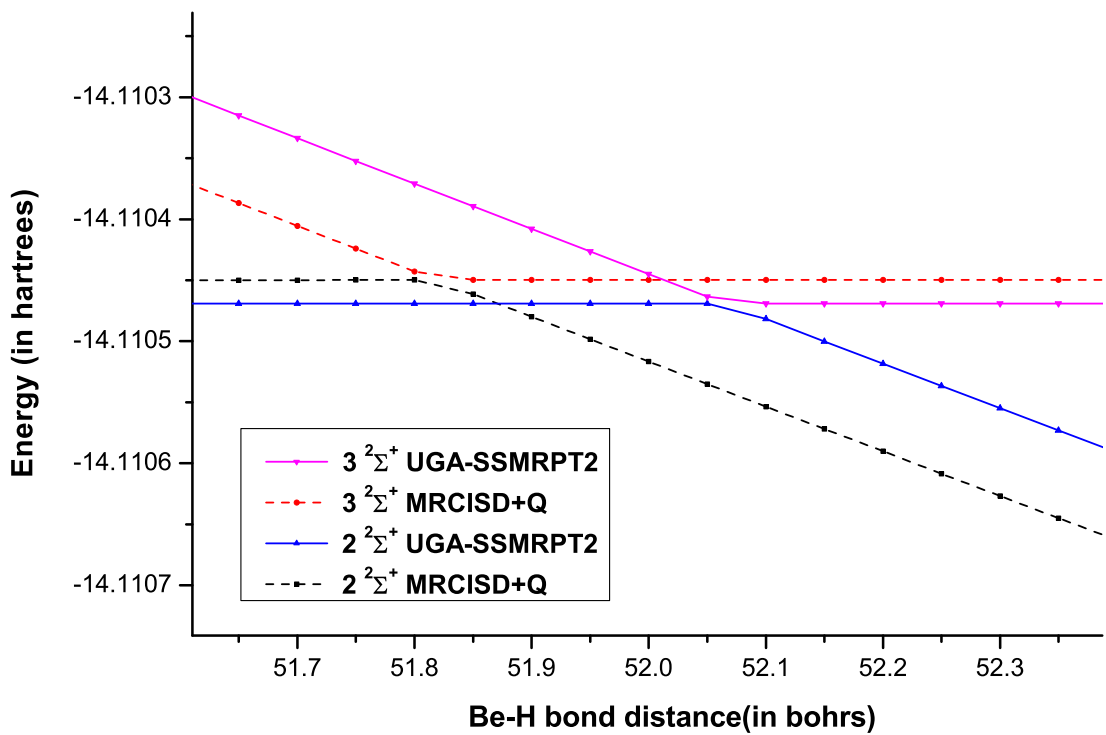}
        \caption{Avoided crossing region between the 1st and 2nd excited $^2\Sigma^+$  states of BeH$^{2+}$ computed using the aug-cc-pVDZ basis}
        \label{BeH2+_2_3_zoomed}
\end{figure}
The MRCISD+Q data points have been shown only in the zoomed area of Fig. \ref{BeH2+_2_3}, that is in Fig. \ref{BeH2+_2_3_zoomed} for clarity of presentation. As is evident from Table \ref{A vs P}, all three doublet states of BeH$^{2+}$ differ by a few 10$^{-5}$ hartrees from the corresponding MRCISD+Q. This is further exemplified by the distance at which the avoided crossing occurs between state 2 and 3 of the manifold for the two methods, as can be seen in Table \ref{Avoided crossing comparison}.\\
\\
\textbf{\underline{Weak ionic-covalent avoided crossings in LiF and BN:}}\\
\\
\textbf{LiF:}\\
The very weak avoided crossing between the ionic-covalent states of LiF has been a major challenge for all SS type of multireference theories. Relaxation of the starting coefficents is crucial due to the significant difference in the contribution of the constituent configurations throughout the dissociation profile, before and after the inclusion of dynamic correlation. It is well known \cite{Pathak2017AApplications} that a state-specific model for a multireference perturbation theory always fails to describe this particular avoided crossing.
\par
Although we have already established the efficacy of UGA-SSMRPT2 in describing very weak avoided crossings in BeF$^{2+}$ and BeH$^{2+}$ in the previous sections, our theory cannot ameliorate the 'double crossing' problem near the weak avoided crossing region between the ground and 1st excited states of LiF. We discuss the rationale behind this observation in the concluding remarks of this sub-section. We employ the cc-pVDZ for Li atom and aug-cc-pVDZ basis for F atom, and the minimal (2,2) CAS has been used for the starting CASSCF function. All orbitals were correlated in the SSMRPT computation.
\begin{figure}[H]
     \centering
        \includegraphics[width=0.75\textwidth]{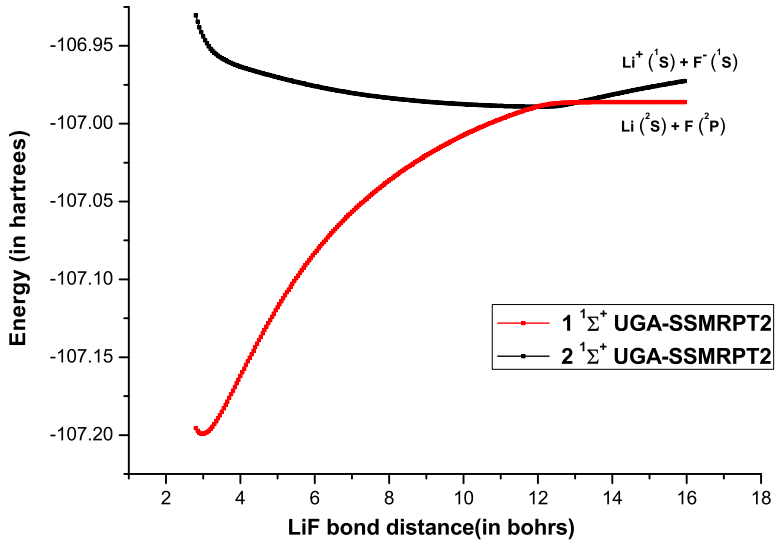}
        \caption{Ground and 1st excited $^1\Sigma^+$  states of LiF computed by UGA-SSMRPT2 using the cc-pVDZ basis (Li) and aug-cc-pVDZ basis (F). See Fig. \ref{LiF_zoomed_mrpt} for a zoomed display of the double crossing between the two states.}
        \label{LiF_mrpt}
\end{figure}
\begin{figure}[H]
     \centering
        \includegraphics[width=\textwidth]{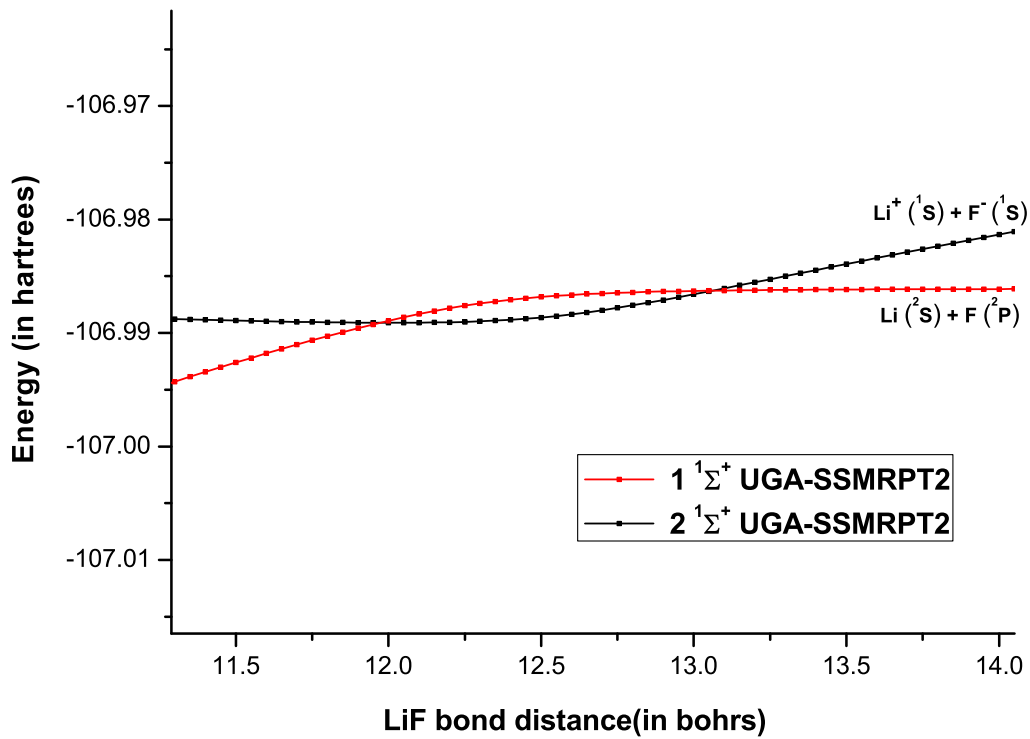}
        \caption{The double crossing between the ground and 1st excited $^1\Sigma^+$  states of LiF computed by UGA-SSMRPT2 using the cc-pVDZ basis (Li) and aug-cc-pVDZ basis (F)}
        \label{LiF_zoomed_mrpt}
\end{figure}
\noindent
\textbf{BN:}\\
We test our theory for another weakly avoided ionic-covalent curve crossing between the 1st and 2nd excited $^3\Pi$ states of BN. It has been established in recent years \cite{Mahmoud2015TheoreticalBN,Bauschlicher1996} that the ground state of BN is $^3\Pi$ and is separated by $\sim$ 200 cm$^{-1}$ from the $^1\Sigma^+$ state. The 1st and 2nd excited $^3\Pi$ states have the same asymptotic limit\cite{Karna1985GroundStudy}, B($^2$P) and N($^2$D), unlike the situation we studied in LiF. The (2)$^3\Pi$ state is formed by excitation of a bonding electron to the non-bonding orbital of B, essentially resulting in a sort of charge transfer from N to B. The (3)$^3\Pi$ state arises from a $\pi$ to $\pi^*$ excitation thus keeping the overall charge of the state neutral.\par
To demonstrate our rationale behind the failure of state-specific perturbation theories in LiF weakly avoided ionic-covalent crossing, we show that the same problem arises when describing the 1st and 2nd excited $^3\Pi$ states in BN.\\ We employ a cc-pVTZ basis and use the full valence (8,8) CAS as our starting function. All orbitals were correlated in our computation.
\begin{figure}[H]
     \centering
        \includegraphics[width=0.70\textwidth]{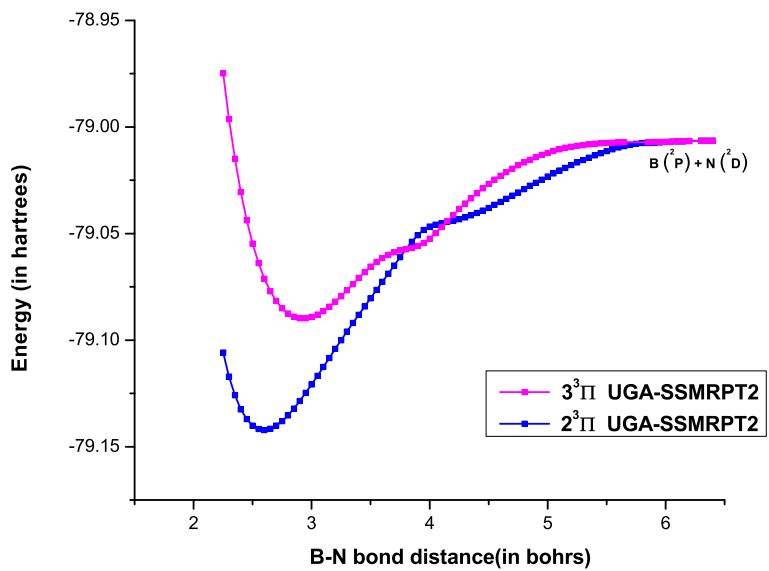}
        \caption{1st and 2nd excited $^3\Pi$  states of BN computed by UGA-SSMRPT2 using the cc-pVTZ basis. See Fig. \ref{BN_mrpt_zoomed} for a zoomed display of the double crossing between the two states.}
        \label{BN_mrpt}
\end{figure}
\noindent
\begin{figure}[H]
     \centering
        \includegraphics[width=0.70\textwidth]{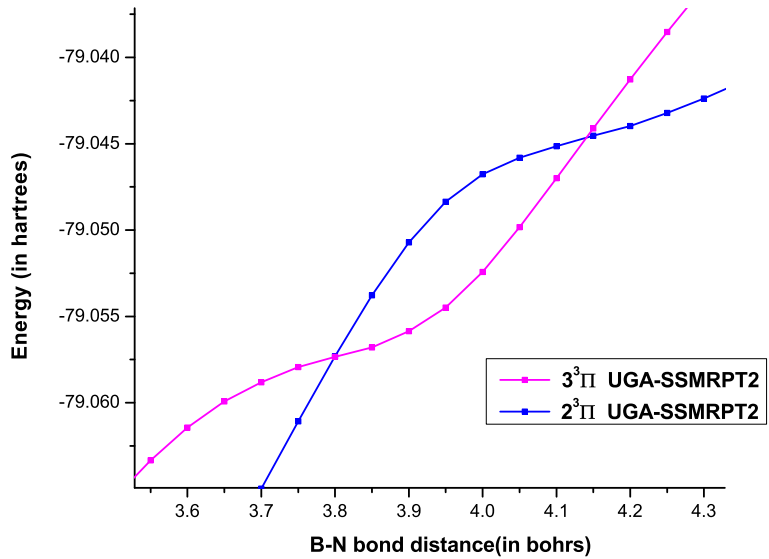}
        \caption{Double crossing region between the 1st and 2nd excited $^3\Pi$  states of BN computed by UGA-SSMRPT2 using the cc-pVTZ basis}
        \label{BN_mrpt_zoomed}
\end{figure}
\noindent
\textbf{Analysis of the double crossing phenomenon:}\\
The inability to reproduce the weakly avoided ionic-covalent crossing in LiF has been encountered by all state-specific PTs \cite{Spiegelmann1984TheCurves,Nakano1993QuasidegenerateFunctions,Malrieu1995MulticonfigurationalProperties,Finley1998TheMethod,Granovsky2011ExtendedTheory,Sharma2016Quasi-degenerateStates,Pathak2017AApplications}. These authors have circumvented the spuriosity, manifested in the form of a double crossing, by constructing an intermediate Hamiltonian consisting both of the same symmetry states, thus forcing the two roots to be dependent on each other. There is no such constraint with a state-specific approach, as the two states are solutions of two different SS-effective Hamiltonians. Nevertheless, as shown in the above applications, our SS theory accurately reproduces the transition between various configurations in a manifold of same symmetry states in most of the systems studied. This includes the very weakly avoided crossings in BeF$^{2+}$ and BeH$^{2+}$.
\par
The failure to describe the weak ionic-covalent avoided crossings in LiF and BN may be attributed to the insufficient description of (i) anionic (N+1 electron) correlation in the asymptotic region and (ii) orbital relaxation effects caused by polarization of the charged species, at a 2nd order perturbative level. If the asymptotes are for differently charged sectors, the insufficient inclusion of dynamical correlation will distort and enhance the differences in gradients of the two states concerned as is the case in LiF. On the other hand, as in BN, the asymptotes of the two states are the same and here the insufficient inclusion of state-specific orbital relaxation is the main contributor to the 'double crossing' artefact.
\par
The polarization effects on the molecular orbitals are best described via the action of one-body cluster operators and its various powers on the spatial orbitals, an effect we would like to call a Thouless-type relaxation \cite{Thouless1960StabilityTheory}. The reason we want to distinguish this with the true Thouless relaxation is because we use a normal ordered exponential of a one body operator to induce the orbital relaxation whereas in the Thouless relaxation a simple exponential of a one-body operator was used. Despite the difference in the structure of the relaxation operator, the physics induced by them is very similar, although our normal ordered ansatz gives a more compact formulation. At the second order of energy, only the linear power of T$_1$ is present in the first order perturbed wavefunction which poorly describes these orbital relaxation effects for a charged species. The combined effect of the insufficient correlation for the anionic species and lack of proper relaxation in the anionic fragment of the molecule results in an artefact manifested via the two same space-spin symmetry potential energy curves crossing each other twice, as a way of joining continuously with their behaviour away from the avoided crossing zone. A lucid analytic demonstration of this 'double crossing' phenomenon using a three state model problem via a low order correlation theory was presented by Spiegelmann and Malrieu\cite{Spiegelmann1984TheCurves}. It is important to mention here that we have already studied the LiF system with the fully non-perturbative UGA-SSMRCC where the weakly avoided crossing was seen to be perfectly reproduced. The results would be demonstrated in a forthcoming publication\cite{TBP}. This bolsters our rationale that an all order state-specific orbital Thouless relaxation is crucial to properly reproduce such ionic-covalent weak avoided crossings and a single-root non-perturbative formalism could be successful to describe this phenomenon.
\par
In contrast, for the systems like BeF$^{2+}$ and BeH$^{2+}$ the two states of interest have orbitals already optimized for a cationic species and thus do not require much orbital relaxation for the different cationic fragments in the two states. Moreover, the dynamical correlation for a cationic (N-1)/(N-2) species is described well even at a perturbative level by the N-electron H$_0$. This allows for a proper description of the very weak avoided crossings in these manifolds of PEC using a 2nd order perturbative correlation theory.
\section{Summary and future outlook}
We have studied the performance of the UGA-SSMRPT2 in the description of various intricate modulations such as interlacing and avoided crossings between PEC of states belonging to the same space-spin symmetry. The accurate depiction of these features in a PEC using the simple and computationally inexpensive perturbative theory, despite the fact that the solutions belong to unrelated state-specific effective Hamiltonians, encourages its extension into the qualitative understanding of bonding and reaction mechanisms of chemically relevant systems.
\par
We have also demonstrated with specific examples the rationale behind the limitations of a state-specific perturbation theory in the cases of weak avoided crossings between ionic and covalent curves. We are hoping to present a rigorously size-extensive, multi-state version of UGA-SSMRPT2 in the near future which should amend this pitfall.
\par
Lastly, we have successfully validated the sufficiency condition we had suggested in our earlier work \cite{Sen2015UnitaryApplications} by implementing the rigorous projection scheme to solve for cluster amplitudes. The performance of both the schemes has been shown to be commensurate with each other, and as such one can safely proceed with the cheaper alternative imparted by our sufficiency. We shall investigate whether the aforementioned equivalence in performance between these two schemes also holds true for its non-perturbative counterpart, the UGA-SSMRCC, in a forthcoming publication.
\section{Supplementary Material}
Please see the supplementary material containing the data points used in construction of the PECs.
\section{Data Availability}
The data that support the findings of this study are available in tabular form in the supplementary material.
\section{Acknowledgement}
DC and RK acknowledge UGC, KH acknowledges his CSIR grant for financial assistance. DM thanks the SN Bose National Centre for Basic Sciences, where he was affiliated during the initial stages of this work, for the SN Bose Chair Professorship funds. We thank Professors Ankan Paul and Satrajit Adhikari for providing laboratory and computational facilities at IACS. The insightful discussions with Dr. Avijit Sen and Dr. Sangita Sen were extremely helpful for starting the programmatic renovation. The authors also thank Professors Aristides Mavridis and Demeter Tzeli for kindly providing the data points for BC used in their papers \cite{Tzeli2001First-principlesAlC-,Tzeli2001Accurate2}.\par\
DM dedicates this paper to the memory of Werner Kutzelnigg, a dear friend and a close collaborator for many years.
%\begin{acknowledgement}

%\end{acknowledgement}
\bibliography{main.bib,molpro.bib}
\end{document}